\documentclass[12pt]{article} 

\usepackage{geometry}
\usepackage{amsmath} % already in sashorthand but should be defined earlier here to use \numberwithin
\numberwithin{equation}{section}
\allowdisplaybreaks % Allows alligned equations to break over pages

\usepackage{color} 
\definecolor{darkblue}{rgb}{0.1,0.1,.7}
\usepackage[colorlinks, linkcolor=darkblue, citecolor=darkblue, urlcolor=darkblue, linktocpage]{hyperref} 

\usepackage[square, comma, sort&compress,numbers]{natbib} % reference management

\usepackage[margin=10pt,font=small,labelfont=bf]{caption}

\usepackage[]{sashorthand} % Provides useful commands and packages. It has as three options:
\newcommand{\dDisct}[1]{\text{dDisc}_{t}\left[#1\right]}
\newcommand{\pre}{\mathsf{p}}
\newcommand{\D}{\mathcal{D}}
\renewcommand{\O}{\cO}
\newcommand{\DD}{\mathrm{D}}

\newcommand{\ope}{\mathcal{O}}
\def\tfo{{}_{2}F_{1}}
\newcommand{\shift}[1]{\Pi_{#1}}

\begin{document}
	\vspace*{-.6in} \thispagestyle{empty}
	\begin{flushright}
	\end{flushright}
	\vspace{.2in} {\Large
		\begin{center}
			{\bf More Analytic Bootstrap: \\ Nonperturbative Effects and Fermions\vspace{.1in}}
		\end{center}
	}
	\vspace{.2in}
	\begin{center}
		{\bf 
			Soner Albayrak$^{a,b}$, David Meltzer$^{a,b}$, David Poland$^{a,b}$}
		\\
		\vspace{.2in} 
		$^a$ {\it  Department of Physics, Yale University, New Haven, CT 06511}\\
		$^b$ {\it  Walter Burke Institute for Theoretical Physics, Caltech, Pasadena, CA 91125}
	\end{center}
	
	\vspace{.2in}
	
	\begin{abstract}

We develop the analytic bootstrap in several directions. First, we discuss the appearance of nonperturbative effects in the Lorentzian inversion formula, which are exponentially suppressed at large spin but important at finite spin. We show that these effects are important for precision applications of the analytic bootstrap in the context of the 3d Ising and O(2) models. In the former they allow us to reproduce the spin-2 stress tensor with error at the $10^{-5}$ level while in the latter requiring that we reproduce the stress tensor allows us to predict the coupling to the leading charge-2 operator. We also extend perturbative calculations in the lightcone bootstrap to fermion 4-point functions in 3d, predicting the leading and subleading asymptotic behavior for the double-twist operators built out of two fermions. 

	\end{abstract}
	
	\newpage
	
	\tableofcontents
	
	\newpage

\section{Introduction}
	\label{sec:intro}

Conformal field theories have numerous applications, ranging from describing continuous phase transitions in condensed matter and statistical systems to describing the observables of quantum gravity in holography. Since the seminal works~\cite{Ferrara:1973yt,Polyakov:1974gs}, the dream of the conformal bootstrap has been to classify and solve the whole landscape of CFTs. Exciting progress towards this goal has been made using both numerical and analytic techniques, as recently reviewed in~\cite{Poland:2018epd}. Some notable numerical results include precision determinations of the leading scaling dimensions and 3-point function coefficients in the 3d Ising and $O(N)$ vector models~\cite{ElShowk:2012ht,Kos:2013tga,Kos:2014bka,Kos:2015mba,Kos:2016ysd}. 

On the analytic side, various attempts have been made to analyze the bootstrap equations in different Lorentzian regimes, including the lightcone and Regge limits. In the former case, it was shown in~\cite{Komargodski:2012ek,Fitzpatrick:2012yx} that the bootstrap applied to scalar 4-point functions $\<\phi\phi\phi\phi\>$ implies the existence of sequences of ``double-twist'' operators $\sim \phi \partial^{\ell} \partial^{2n} \phi$ at large spin $\ell \gg 1$, whose twists asymptotically approach $\tau \sim 2 \tau_{\phi} + 2n$, with corrections determined by the low twist operators appearing in the $\phi \times \phi$ OPE after the identity operator.\footnote{See also \cite{Alday:2007mf} for an earlier derivation of this result.} A variety of studies have since extended these calculations to higher orders in the large-spin expansion~\cite{Alday:2015ewa,Alday:2016njk,Simmons-Duffin:2016wlq}, as well as to some situations with external currents and stress tensors~\cite{Li:2015itl,Hofman:2016awc,Sleight:2018epi,Sleight:2018ryu}.

Instead going to the Regge limit allows one to probe the leading Regge trajectories of the theory, whose resummations can be captured by conformal Regge theory~\cite{Costa:2012cb}. In large $N$ CFTs the bootstrap equations in this limit relate these contributions to double-twist operators with $n \sim \ell\gg1$~\cite{Li:2017lmh} and further constraints from unitarity and causality imply a number of nontrivial positivity conditions on how light operators couple to the leading Regge trajectory~\cite{Afkhami-Jeddi:2016ntf,KPZ2017,Costa:2017twz,Meltzer:2017rtf,Afkhami-Jeddi:2017rmx,Afkhami-Jeddi:2018own,Afkhami-Jeddi:2018apj}.

An important advance was made in~\cite{Caron-Huot:2017vep}, which developed a Lorentzian inversion formula for CFT correlators. Among other things, this formula shows that the families of operators described above are analytic in spin and gives a formalism for extending their large-spin expansions down to finite values of the spin. Some of these implications were recently worked out in~\cite{Simmons-Duffin:2017nub,Kravchuk:2018htv}, which make manifest the relation between the Lorentzian inversion formula and light-ray operators. These developments are also closely connected to an improved understanding of crossing kernels and 6j symbols of the conformal group~\cite{Sleight:2018epi,Sleight:2018ryu,Cardona:2018dov,Cardona:2018qrt,Liu:2018jhs}.

In the present work we take some modest steps in further advancing these analytic approaches to the bootstrap. We start in section~\ref{sec:scalar} by reviewing the lightcone bootstrap and Lorentzian inversion formula applied to the leading twist trajectories in scalar 4-point functions, emphasizing effects which are nonperturbative in the large-spin expansion but important at finite spin. Our main observation is that these effects are important for precision analytic predictions of the leading twist spectra in the 3d Ising and O(2) models. In particular, by extending the leading twist family of the 3d Ising model down to spin 2, including these effects and inputting known data from the numerical bootstrap, one can reproduce the twist of the stress tensor with error at the $10^{-5}$ level, improving on the previous lightcone bootstrap analysis of~\cite{Simmons-Duffin:2016wlq}. In the O(2) model there is a larger error on the inputs, particularly in the dimension and coupling to the leading charge-2 operator, but by requiring that the leading charge-0 trajectory contains the stress tensor, we obtain a prediction for this coupling as well as a picture of the leading twist spectra.

In section~\ref{sec:fermion} we extend the analytic bootstrap in another direction, namely to 4-point functions of fermions in 3d. A similar analysis at leading order in the large-spin expansion was carried out in 4d in~\cite{Elkhidir:2017iov}. In this work we carry out the large-spin expansions for the leading twist trajectories and their OPE coefficients to subleading order, in the hope that it will provide a valuable cross-check to future derivations of the Lorentzian inversion formalism for 3d fermions and can be eventually applied to interesting 3d CFTs with fermions. This is in part motivated by recent progress in the numerical bootstrap for such theories, including the 3d Gross-Neveu-Yukawa models~\cite{Iliesiu:2015qra,Iliesiu:2017nrv} and the minimal 3d $\mathcal{N}=1$ SCFT~\cite{Rong:2018okz,Atanasov:2018kqw}.\footnote{See also~\cite{Karateev:2019pvw} for recent progress in 4d.} We conclude with a discussion of future directions in section~\ref{sec:discussion} and give a brief review of the embedding formalism and other technical formulas in appendices~\ref{sec:EmbeddingReview} and~\ref{sec:ExpansionOfS}. 

\section{Scalar Analytic Bootstrap}
\label{sec:scalar}

We will begin our discussion with a review of the analytic bootstrap for scalar 4-point functions $\<\phi_1\phi_2\phi_3\phi_4\>$, which take the general form
\be
\<\phi_1(x_1)\phi_2(x_2)\phi_3(x_3)\phi_4(x_4)\> = \left(\frac{x_{24}}{x_{14}}\right)^{\Delta_{12}}\left(\frac{x_{14}}{x_{13}}\right)^{\Delta_{34}}\frac{G(z,\bar{z})}{x_{12}^{\Delta_1+\Delta_2}x_{34}^{\Delta_3+\Delta_4}}\;,
\ee
where $x_{ij} = x_i - x_j$ and $\Delta_{ij} = \Delta_i - \Delta_j$, with $\Delta_i$ the scaling dimension of $\phi_i$. The conformal cross-ratios $(z,\bar{z})$ are given by
\begin{equation}
z\zb=\frac{x_{12}^2x_{34}^2}{x_{13}^2x_{24}^2}\quad,\quad (1-z)(1-\zb)=\frac{x_{14}^2x_{23}^2}{x_{13}^2x_{24}^2}\;.
\end{equation} 
The function $G(z,\bar{z})$ can be expanded in conformal blocks for the $\f_{1}\f_{2}\rightarrow\f_{3}\f_{4}$ OPE as
\be
G(z,\bar{z}) = \sum_{\cO} f_{12\cO} f_{43\cO} g_{h,\bar{h}}^{r,s}(z,\bar{z})\;,
\ee
where we have introduced the variables
\begin{subequations}
\begin{alignat}{2}
r&=\frac{\Delta_{1}-\Delta_{2}}{2}\;,\qquad   &&s=\frac{\Delta_{3}-\Delta_{4}}{2}\;,
\\
h&=\frac{\Delta-\ell}{2}\;,  
&&\bar{h}=\frac{\Delta+\ell}{2}\;,
\end{alignat}
\end{subequations}
where $\Delta$ and $\ell$ are the scaling dimension and spin of the exchanged operator. This parameterization is convenient for the lightcone expansion, where $\hb$ is the natural expansion parameter.

\subsection{Scalar conformal blocks}
In general dimensions the conformal blocks are not known in a simple closed form,\footnote{See~\cite{Poland:2018epd} for a review of various methods to calculate the blocks.} but in the limit where two operators become light-like separated they display a universal behavior which makes analytic results possible. In terms of the conformal cross-ratios, if we take the limit $z \rightarrow 0$, then the leading order behavior of conformal blocks in any dimension is
\begin{equation}
g_{h,\hb}^{r,s}(z,\zb)\underset{z\rightarrow 0}{\simeq} z^h \zb^{\hb}
\pFq{2}{1}{\hb-r,\hb+s}{2\hb}{\zb}\;.
\label{2.39 of Dolan:2011dv}
\end{equation}
This approximation is sufficient to compute the leading large-$\ell$ corrections to the spectrum of double-twist operators using analytic bootstrap techniques, as was first done in~\cite{Komargodski:2012ek,Fitzpatrick:2012yx}.

In this paper we will be interested in higher order corrections in the large-$\ell$ expansion~\cite{Alday:2015ewa,Alday:2016njk,Simmons-Duffin:2016wlq}, hence we will need subleading terms in the above expansion. In $d=3$ dimensions, which will be the focus of our paper, one can use dimensional reduction to expand the conformal block in terms of 2d conformal blocks~\cite{Hogervorst:2016hal}, or equivalently use an $\SL(2,\mathbb{R})$ expansion~\cite{Simmons-Duffin:2016wlq}. These two methods are equivalent since the 2d blocks are a simple combination of 1d, or $\SL(2,\mathbb{R})$, blocks.

The conformal blocks in any dimension can be expanded as\footnote{In \cite{Simmons-Duffin:2016wlq}, the expansion is given for an $\SO(d,2)$ block $G$ symmetric under $r$ and $s$ exchange. We use the more conventional conformal block $g$, which relates to $G$ there as $g_{h,\hb}^{\text{here},r,s}(z,\zb)=((1-z)(1-\zb))^rG_{h,\hb}^{r,s}(z,\zb)$.}
\bea\label{eq:SL2R}
g_{h,\hb}^{r,s}(z,\zb) &=\sum\limits_{n=0}^\infty\sum\limits_{j=-n}^n A_{n,j}^{r,s}(h,\hb)z^{h+n} k^{r,s}_{2(\bar{h}+j)}(\bar{z})\;,
\\
k^{r,s}_{2\bar{h}}(\bar{z})&=\zb^{\hb}\pFq{2}{1}{\hb-r,\hb+s}{2\hb}{\zb}\;,
\eea 
where $k^{r,s}_{2\bar{h}}(z)$ is the $\SL(2,\mathbb{R})$ block. Using the decomposition mentioned above, or by solving the Casimir differential equation, the first two levels are straightforward to work out and are given by
\bea 
A^{r,s}_{0,0}(h,\hb)&=1\;,\\
A^{r,s}_{1,-1}(h,\hb)&=\frac{h-\bar{h}}{2 h-2 \bar{h}+1}\;,\\
A^{r,s}_{1,0}(h,\hb)&=\frac{1}{2} \left(\frac{r s \left(h-2 \bar{h}^2+2 \bar{h}-1\right)}{(2 h-1) (\bar{h}-1) \bar{h}}-h-r+s\right)\;,\\
A^{r,s}_{1,1}(h,\hb)&=\frac{(h+\bar{h}-1) (\bar{h}-r) (\bar{h}+r) (\bar{h}-s) (\bar{h}+s)}{4 \bar{h}^2 (2 \bar{h}-1) (2 \bar{h}+1) (2 h+2 \bar{h}-1)}\;.
\eea

\subsection{Lightcone bootstrap review} 
\label{sec:LightconeBootstrapReview}

In the following calculations, we review the work of \cite{Simmons-Duffin:2016wlq} which solves the lightcone bootstrap in a perturbative expansion. We start by considering the 4-point function of identical scalars, $\<\f\f\f\f\>$. The 4-point function is invariant under $1\leftrightarrow3$ (or $s\leftrightarrow t$) crossing, which implies
\begin{equation}
\left(\frac{(1-z)(1-\bar{z})}{z\zb}\right)^{2h_{\f}}\sum_{\ope}P_{\f\f\ope}g_{h_\ope ,\hb_\ope}(z,\zb)=\sum_{\ope}P_{\f\f\ope}g_{h_\ope ,\hb_\ope}(1-\zb,1-z)\;, \label{eq:4sCross}
\end{equation}
where $P_{\f\f \ope}=f_{\f\f\O}^{2}$.

We can now consider this equation in the lightcone limit $z\ll1-\bar{z}\ll1$. In the limit $z\ll1$, the identity operator dominates on the left-hand side, while taking $1-\bar{z}\ll1$ allows us to use the $\SL(2,\mathbb{R})$ expansion on the right-hand side. In this limit, the crossing equation becomes
\begin{equation}
\left(\frac{1-\bar{z}}{z}\right)^{2h_{\f}}\approx\sum_{\ope}P_{\f\f\ope}(1-\bar{z})^{h_{\ope}}k_{2\hb_\ope}(1-z)\;.
\end{equation}

By the arguments of \cite{Fitzpatrick:2012yx,Komargodski:2012ek}, in order to match the $z\rightarrow0$ divergence on the left-hand side, which is not present in any individual $t$-channel block, we need to sum over operators with unbounded spin on the right-hand side. Specifically, we need a tower of ``double-twist'' operators in the $t$-channel, $[\f\f]_{0,\ell}$, such that $h_{0,\ell}\rightarrow 2h_{\f}$ as $\ell\rightarrow\infty$. 

At this point we could use a Bessel function approximation of the blocks to derive the large-$\ell$ asymptotics of the OPE coefficients, but it will be more useful to use our knowledge of $1d$ generalized free field theories to write down the exact sum \cite{Simmons-Duffin:2016wlq}
\begin{subequations}
\label{eq:SL2R_Sum}
\be 
\sum_{\substack{\bar{h}=-a+l \\ \ell=0,1,...}}S_{a}(\bar{h})k_{2\bar{h}}(1-z)=\left(\frac{z}{1-z}\right)^{a}\;,
\ee 
where
\be 
S_{a}(\bar{h})=\frac{\Gamma(\bar{h})^{2}\Gamma(\bar{h}-a-1)}{\Gamma(-a)^{2}\Gamma(2\bar{h}-1)\Gamma(\bar{h}+a+1)}\;.
\ee 
\end{subequations}

From (\ref{eq:SL2R_Sum}), we now have at large $\ell$, where $\bar{h}\approx 2h_{\f}+\ell$, the following result for the OPE coefficients:
\begin{equation}
P_{\f\f\ope}(\bar{h})\sim S_{-2h_{\f}}(\bar{h})\;.
\end{equation}
The ``$\sim$'' is because with this approach we can only find the asymptotic expansion for the OPE coefficients at large $\bar{h}$. Note that by expanding to higher orders in $1-\bar{z}$ one can also prove the existence of operators $[\f\f]_{n,\ell}$ which have $h_{n,\ell}\rightarrow 2h_{\f}+n$ as $\ell\rightarrow\infty$. 

To extend these calculations to higher orders in the large-$\bar{h}$ expansion, we can use the $\SL(2,\mathbb{R})$ expansion on both sides of the crossing equation (\ref{eq:4sCross}), expand in $z\ll1-\bar{z}\ll1$, and then use (\ref{eq:SL2R_Sum}) to unambiguously match generic powers of $z$ in the $s$-channel to the large-spin asymptotics of double-twist operators in the $t$-channel.

Of course, there are subtleties in this procedure which we have glossed over. First, for the arguments of \cite{Fitzpatrick:2012yx,Komargodski:2012ek} to work when matching a power of $z$ in the $s$-channel to an infinite sum of $t$-channel blocks, we need the $s$-channel term to be more divergent than any individual $t$-channel block. From (\ref{eq:4sCross}), we see this is only true if $h_{\ope}<2h_{\f}$. However, as noted in \cite{Alday:2015ewa}, we can make any generic, individual power of $z^{a}$ on the left-hand side of (\ref{eq:4sCross}) as divergent as we like by repeatedly acting with a $\SL(2,\mathbb{R})$ Casimir differential operator:
\be
\mathcal{C}\equiv (1-z)^{2}z\partial_{z}^{2}+(1-z)^{2}\partial_{z}\;.
\ee
Since the $t$-channel $\SL(2,\mathbb{R})$ blocks are eigenfunctions of this Casimir, these differential operators leave the form of the $t$-channel expansion unchanged. Therefore, by acting with this differential operator sufficiently many times we can make the $s$-channel more divergent than the crossed channels.

In the terminology of \cite{Simmons-Duffin:2016wlq}, generic powers $z^{a}$ are ``Casimir-singular''. On the other hand, terms like $z^{n}$ and $z^{n}\log(z)$, with $n$ a non-negative integer, are called ``Casimir-regular''. If we repeatedly act with $\mathcal{C}$ on these terms, we eventually get $0$. Therefore, we cannot use the arguments of \cite{Fitzpatrick:2012yx,Komargodski:2012ek} to match these terms with large-spin asymptotics of double-twist operators and they are more sensitive to finite-spin effects.

In the study of the lightcone bootstrap, there are a few places where Casimir-regular terms can appear. The first is if we start the $\SL(2,\mathbb{R})$ sum (\ref{eq:SL2R_Sum}) at a generic point $\bar{h}_{0}$:
\begin{equation}
\sum_{\substack{\bar{h}=\bar{h}_{0}+\ell \\ \ell=0,1,...}}S_{a}(\bar{h})k_{2\bar{h}}(1-z)=\left(\frac{z}{1-z}\right)^{a}+\mathcal{A}(\bar{h}_0)\;,
\end{equation}
where $\mathcal{A}(\bar{h}_0)$ is defined in \cite{Simmons-Duffin:2016wlq} (we will not need its explicit form). Since the choice of starting point only affects a finite number of blocks, changing the lower limit of the sum will not affect predictions for large-spin asymptotics. 

Another important issue is that our sums over blocks are not actually integer spaced. In general, the double-twist operators will get anomalous dimensions which also depend on the spin. Therefore, for a tower of double-twist operators $\O_{\ell}$ parametrized by the spin $\ell$ we have 
\begin{equation}
\bar{h}_{\O_{\ell}}=2 h_{\phi}+\ell+\delta h_{\O_{\ell}}\;,
\end{equation}
where $\delta h_{\O_\ell}$ is half the anomalous dimension with respect to the generalized free field value.

To account for this effect, we can reparametrize our sum by inserting a Jacobian:
\begin{equation}
\sum\limits_{\ell=0}^{\infty}\frac{\partial \bar{h}_{\O_\ell}}{\partial \ell} P(\bar{h}_{\O_\ell})k_{2\bar{h}_{\O_\ell}}(1-z)=\sum\limits_{\ell=0}^{\infty}P(2h_{\phi}+\ell)k_{4h_{\phi}+2\ell}(1-z)+\ldots\;,
\end{equation}
where the dropped terms are Casimir-regular in $z$. In the current discussion we will not be concerned with matching Casimir-regular terms, but will rather focus on how we can use the $\SL(2,\mathbb{R})$ expansion to match individual conformal blocks in the $s$-channel.

Thus, let us now consider the effect of single generic conformal block $g_{h_i,\bar{h}_i}(z,\bar{z})$ in the s-channel. In the limit $z\ll1-\bar{z}\ll1$ it has the general form:
\be
g_{h_i,\bar{h}_i}(z,\bar{z})=\left(\frac{z}{1-z}\right)^{h_i}\left(A_i \log(1-\bar{z})+B_i+\O(1-\bar{z})\right)+\ldots\;,
\ee
where we have only written the leading order terms. It is straightforward to include $\SL(2,\mathbb{R})$ descendants using the results of \cite{Simmons-Duffin:2016wlq,Hogervorst:2016hal}, and to expand to higher orders in $(1-\bar{z})$ using the explicit form of the hypergeometric functions, where such terms are needed to fix the corrections for higher twist towers, i.e. $[\phi\phi]_{n>0}$.\footnote{We will drop the label $\ell$ when referring to a given twist tower.} Here we will be primarily interested in the form of the correction for the leading twist $[\f\f]_{0}$ operators.

The crossing equation then becomes
\begin{multline}
\left(\frac{z}{1-z}\right)^{-2h_\phi}+\sum_i \left(\frac{z}{1-z}\right)^{h_i-2h_\phi}(A_i\log(1-\bar{z})+B_i+\order{1-\bar{z}})\\=\sum\limits_{\ope\in[\phi\phi]_0}P_{\phi\phi\ope}(1-\bar{z})^{h_\ope-2h_\phi}k_{2\bar{h}_\ope}(1-z)+\ldots.\label{eq:CrossingForFourScalars}
\end{multline}

To match the $\log$ terms we have to expand in the anomalous dimension, $h_{\ope}=2h_{\f}+\delta h_\ope$, and we find
\bea[5] 
P_{\phi\phi\ope}\sim{}&2\frac{\partial \bar{h}_\ope}{\partial l}\left[S_{-2h_\phi}(\bar{h}_\ope)+\sum_{i} B_iS_{h_i-2h_\phi}(\bar{h}_\ope)\right]\;,\label{eq:asymptoticP}\\
\delta h_\ope P_{\phi\phi\ope}\sim{}&2\frac{\partial \bar{h}_\ope}{\partial l}\left[\sum_{i} A_iS_{h_i-2h_\phi}(\bar{h}_\ope)\right]\;.
\eea 
The factors of $2$ are because we only sum over double-twist operators  of even spin in the $t$-channel. To find the asymptotic, large-spin behavior of the anomalous dimensions we then just need the ratio of these terms:
\begin{equation}
\delta h_\ope \sim \frac{\sum_{i} A_iS_{h_i-2h_\phi}(\bar{h}_\ope)}{S_{-2h_\phi}(\bar{h}_\ope)+\sum_{i} B_iS_{h_i-2h_\phi}(\bar{h}_\ope)}\;.\label{eq:asymptoticdh}
\end{equation}
In order to compute $P_{\phi\phi\ope}$ we can then plug this into the relation $\frac{\partial \bar{h}_\ope}{\partial l}=\left(1-\frac{\partial \delta h_\ope}{\partial \bar{h}_\ope}\right)^{-1}$.

\subsection{Inversion formula approach}
\label{sec:inversion}

An alternative elegant way to calculate OPE data is by making use of the Lorentzian inversion formula \cite{Caron-Huot:2017vep,Simmons-Duffin:2017nub}. In addition to providing a resummation of the $1/\ell$ expansion, this formalism also allows one to compute nontrivial nonperturbative effects which are exponentially suppressed at large $\ell$ but may be important at smaller values of $\ell$~\cite{Sleight:2018epi, Cardona:2018dov, Sleight:2018ryu, Liu:2018jhs, Cardona:2018qrt}. Such effects are in fact needed in order to obtain a resummation which is analytic in $\ell$. We would like to take the opportunity to review a derivation of these effects, generalizing previous computations to different external dimensions and arbitrary $\SL(2,\mathbb{R})$ blocks, and also to illustrate their importance in 3d CFTs.

For a 4-point function of scalars $\<\f_1\f_2\f_3\f_4\>$, the CFT inversion formula gives the OPE data for the $s$-channel in terms of two integrals of the function $g(z,\bar{z})$:
\begin{equation}
c(h,\bar{h})=c^{t}(h,\bar{h})+(-1)^{\hb-h}c^{u}(h,\bar{h})\;,
\end{equation}
where
\bea
c^{t}(h,\bar{h})=&\frac{\kappa_{2\bar{h}}}{4}\int\limits_{0}^{1}dzd\bar{z}\mu(z,\bar{z})g^{r,s}_{d-1-h,\bar{h}}(z,\bar{z})\text{dDisc}_{t}[g(z,\bar{z})]\;,\label{eq:ctInv}
\\
\kappa_{2\hb}\equiv&\frac{\Gamma(\hb+r)\Gamma(\hb-r)\Gamma(\hb+s)\Gamma(\hb-s)}{2\pi^2\Gamma(2\hb-1)\Gamma(2\hb)}\;,
\\
\mu(z,\bar{z})=&\bigg|\frac{z-\bar{z}}{z\bar{z}}\bigg|^{d-2}\frac{\left((1-z)(1-\bar{z})\right)^{s-r}}{(z\bar{z})^{2}}\;,
\eea
and we recall that $r=h_{1}-h_2$ and $s=h_3-h_4$. The double discontinuity around $z=0$, which we call the $s$-channel dDisc, is defined by
\be
\text{dDisc}_{s}[g(z,\bar{z})]=\cos\left(\pi(s-r)\right)g(z,\bar{z})-\frac{1}{2}e^{i\pi(s-r)}g(ze^{2\pi i},\bar{z})-\frac{1}{2}e^{i\pi(r-s)}g(ze^{-2\pi i},\bar{z})\;. \label{eq:dDisc}
\ee
The $t$ and $u$-channel double discontinuities are defined in the same way, except around $z=1$ and $z=\infty$ respectively. The term $c^{u}(h,\bar{h})$ is also defined in the same way as (\ref{eq:ctInv}), but with the integration being taken from $-\infty$ to $0$.

In \cite{Caron-Huot:2017vep} the inversion formula was written in terms of $(\Delta,\ell)$, in which case the OPE coefficients for generic $\Delta$ are given by
\begin{equation}
f_{12\O}f_{34\O}=-\Res_{\Delta'=\Delta}c(\Delta',\ell)\;, \quad \text{for fixed $\ell$}.
\end{equation}
Here we need to take residues of $c(h,\bar{h})$ with respect to $h$ at fixed $\bar{h}-h=\ell$, which will introduce some extra Jacobians as in the lightcone bootstrap. We will focus on $c^{t}$ since the $u$-channel can always be found by taking $1\leftrightarrow3$ and multiplying by $(-1)^{\ell}$.

It is convenient to define a generating function for the poles of $c^{t}(h,\bar{h})$:
\be
c^{t}(h,\bar{h})\bigg|_{\text{poles}}=\int\limits_{0}^{1}\frac{dz}{2z}z^{-h}C^{t}(z,\bar{h})\;.
\ee
The outer integral turns powers of $z$ inside $C^{t}(z,\bar{h})$ into poles for $h$. Since we are interested in the low-twist data, and in particular the $n=0$ double-twist operators, we study the small $z$ limit of $C(z,\bar{h})$:
\be
C^{t}(z,\bar{h})\approx \int\limits_{0}^{1}d\bar{z}\frac{(1-\bar{z})^{s-r}}{\zb^{2}}\kappa_{2\bar{h}}k^{r,s}_{2\bar{h}}(\bar{z})\text{dDisc}_{t}\left[\frac{(z\bar{z})^{h_1+h_2}g(1-z,1-\bar{z})}{[(1-z)(1-\bar{z})]^{h_{2}+h_{3}}}\right]\;,
\ee
where we have used crossing symmetry inside the dDisc.
This is a generating function for the $\SL(2,\mathbb{R})$ primaries with respect to $\bar{z}$ and to subtract descendants along $z$ we have to expand the inverted block in (\ref{eq:ctInv}) in powers of $z$. 

One nice example is to consider a 4-point function of identical scalars, $\<\f\f\f\f\>$. As in the lightcone bootstrap, terms regular and logarithmic in $z$ in $C^{t}(z,\bar{h})$ will correspond to corrections of OPE coefficients and scaling dimensions of the double-twist towers $[\f\f]_{n}$, respectively. To see this we will assume the anomalous dimensions of double-twist operators are small and expand $C^{t}(z,\bar{h})$ both in $z$ and the anomalous dimension:
\be
C^{t}(z,\bar{h})\approx  z^{2h_\f} P_{[\f\f]_0}(\bar{h})(1+\delta h_{[\f\f]_0}(\bar{h})\log(z)) + \ldots\;,
\ee
where the $\log(z)$ comes from a single $t-$channel conformal block. Integrating over $z$ we see the term regular in $z$ becomes a single pole while the term logarithmic in $z$ becomes a double pole. Some of these corrections at finite spin were recently derived in the works~\cite{Sleight:2018epi, Cardona:2018dov, Sleight:2018ryu, Liu:2018jhs, Cardona:2018qrt}. We will review these results and present some generalizations.

As an example, we can consider the exchange of a scalar operator $\mathcal{O}$ of twist $\tau_\O=2h_\O$ in the $t$-channel and use the inversion formula to extract the anomalous dimension of the $[\f\f]_0$ tower. In the limit $z\rightarrow 0$, the $\log(z)$ piece of the scalar block $g_{h_\O,h_\O}(1-z,1-\bar{z})$ is known in closed form. In a general 4-point function $\<\f_{1}\f_{2}\f_{3}\f_{4}\>$ the $t$-channel blocks develop logs when $h_{1}+h_{2}=h_{3}+h_{4}$, with the coefficient given by\footnote{The conformal block normalization is the same as in Eq.~(\ref{2.39 of Dolan:2011dv}).}
\begin{multline}
g_{h_\O,h_\O}(1-z,1-\bar{z})\bigg|_{\log(z)}=-\log(z)\frac{\Gamma (2 h_{\O})}{\Gamma (h_{1}-h_{4}+h_{\O}) \Gamma (-h_{1}+h_{4}+h_{\O})}  \\ 
\times (1-\bar{z})^{h_{\O}} \pFq{2}{1}{h_{1}-h_{4}+h_{\O},-h_{1}+h_{4}+h_{\O}}{2 h_{\O}-\frac{d-2}{2}}{1-\bar{z}}\;.
\end{multline}

Focusing for now on the case where the external scalars are the same, we can match the $\log(z)$ term in the generating function, yielding the correction
\begin{multline}
\hspace{-.5cm}(\delta h P)_{[\phi\phi]_0}(\hb)=- f_{\f\f\O}^{2}\frac{\Gamma(2h_\O)}{\Gamma(h_\O)^{2}} \kappa_{2\bar{h}}\int\limits_{0}^{1}\frac{d\bar{z}}{\bar{z}^{2}} k_{2\bar{h}}(\bar{z})\\\times\dDisct{\left(\frac{\bar{z}}{1-\bar{z}}\right)^{2h_{\f}}(1-\bar{z})^{h_\O}\pFq{2}{1}{h_\O,h_\O}{2h_\O-\frac{d-2}{2}}{1-\bar{z}}}\;.
\end{multline}
Notice that this formula gives the product $\delta h \times P $ and one still needs to compute the corrected OPE coefficients, as well as add the $u$-channel contribution (identical up to a factor $(-1)^{\bar{h}-h}$), in order to find the anomalous dimension.

Using a hypergeometric transformation, we can rewrite this as
\begin{multline}\label{eq:zbintegral}
(\delta h P)_{[\phi\phi]_0}(\hb)= - f_{\f\f\O}^{2} \frac{\Gamma(2h_\O)}{\Gamma(h_\O)^{2}}\kappa_{2\bar{h}}  \int_0^1 \frac{d\bar{z}}{\bar{z}^2} \left(\frac{1-\bar{z}}{\bar{z}}\right)^{-\hb}
\pFq{2}{1}{\hb,\hb}{2\hb}{-\frac{\bar{z}}{1-\bar{z}}}
\\
\times \bar{z}^{-h_\O}
\pFq{2}{1}{h_\O,h_\O-\frac{d-2}{2}}{2h_\O-\frac{d-2}{2}}{-\frac{1-\bar{z}}{\bar{z}}}  \dDisct{(1-\bar{z})^{h_\O} \left(\frac{\bar{z}}{1-\bar{z}}\right)^{\Delta_{\phi}}}\;.
\end{multline}
Finally, we can write the hypergeometric functions as a Mellin-Barnes integral and perform the $\bar{z}$ integral using the identity\footnote{We thank David Simmons-Duffin for discussions on these integrals.}
\be
\int_0^1 \frac{d\bar{z}}{\bar{z}(1-\bar{z})} \left(\frac{\bar{z}}{1-\bar{z}}\right)^{\alpha} = 2\pi \delta(i\alpha) \label{eq:zIntegration}
\ee
to find:
\be
(\delta h P)_{[\phi\phi]_0}(\hb)=&-f_{\f\f\O}^2    \sin^2 \left(\pi(h_\O-2h_{\phi})\right) \frac{\Gamma(2h_\O)\Gamma\left(2h_\O-\frac{d-2}{2}\right)}{\pi^2\Gamma(h_\O)^3\Gamma\left(h_\O-\frac{d-2}{2}\right)}   \frac{\Gamma(\hb)^2}{\Gamma(2\hb-1)}
\\&
\times
\frac{1}{2\pi i} \int_{-i\infty}^{i\infty} ds'
\Bigg( \frac{\Gamma\left(h_\O-2h_{\phi}+1+s'\right)^2\Gamma\left(\hb-h_\O+2h_{\phi}-1-s'\right)}{\Gamma\left(\hb+h_\O-2h_{\phi}+1+s'\right)}
\\&\hspace{1.25in} \frac{\Gamma\left(h_\O+s'\right)\Gamma\left(h_\O-\frac{d-2}{2}+s'\right)\Gamma(-s')}{\Gamma\left(2h_\O-\frac{d-2}{2}+s'\right)}\Bigg)\;.
\ee
Finally, summing over the poles of $\Gamma(-s')$ gives
\begin{multline}
(\delta h P)_{[\phi\phi]_0}(\hb)\big|_{\text{pert}}
=-f_{\f\f\O}^{2} \sin ^2(\pi  (h_{\O}-2 h_{\f}))  \frac{\Gamma (2 h_{\O})   \Gamma (h_{\O}-2 h_{\f}+1)^2}{ \pi^2  \Gamma (h_{\O})^2}
\\  \times  \frac{\Gamma (\bar{h})^2 \Gamma(\bar{h}-h_{\O}+2h_{\phi}-1)}{\Gamma (2 \bar{h}-1) \Gamma(\bar{h}+h_{\O}-2h_{\phi}+1)} \pFq{4}{3}{h_{\O},h_{\O}-\frac{d-2}{2},h_{\O}-2 h_{\f}+1,h_{\O}-2 h_{\f}+1}{-\bar{h}+h_{\O}-2 h_{\f}+2,\bar{h}+h_{\O}-2 h_{\f}+1,2 h_{\O}-\frac{d-2}{2}}{1}\;,
\end{multline}
which when expanded at large $\bar{h}$ reproduces the perturbative $1/\bar{h}$ expansion from the lightcone bootstrap. It can also be obtained by expanding the hypergeometric function inside Eq.~(\ref{eq:zbintegral}) at small $\bar{z}$, performing the integrals term-by-term, and resumming the result. However, this resummation contains spurious poles in $\bar{h}$, leading to a non-analytic function, connected to the fact that performing the small $\bar{z}$ expansion inside the integral fails to correctly capture its behavior near $\bar{z} \sim 1$.  

Additionally summing over the poles of $\Gamma\left(\bar{h} -h_\O+2h_\phi-1-s'\right)$ gives the correction
\be
(\delta h P)_{[\phi\phi]_0}(\bar{h})\big|_{\text{nonpert}}
=&- f_{\f\f\O}^2    \sin^2 \left(\pi(h_\O-2h_{\phi})\right) \frac{\Gamma(2h_\O)\Gamma\left(2h_\O-\frac{d-2}{2}\right)}{\pi^2\Gamma(h_\O)^3\Gamma\left(h_\O-\frac{d-2}{2}\right)}   \frac{\Gamma(\bar{h})^4 }{\Gamma(2\bar{h}-1)\Gamma(2\bar{h})}
\\
&\times \frac{ \Gamma(-\bar{h}+h_\O-2h_{\phi}+1)\Gamma(\bar{h}+2h_{\phi}-1)\Gamma(\bar{h}+2h_{\phi}-\frac{d}{2})}{\Gamma(\bar{h}+h_\O+2h_{\phi}-\frac{d}{2})}  \\&\x\pFq{4}{3}{\bar{h}, \bar{h}, \bar{h} + 2 h_{\f}-1,\bar{h}+2 h_{\f}-\frac{d}{2}}{2 \bar{h},\bar{h}-h_{\O}+2 h_{\f},\bar{h}+h_{\O}+2 h_{\f}-\frac{d}{2}}{1}\;.
\ee

The full sum $(\delta h P)_{[\phi\phi]_0}(\bar{h})=(\delta h P)_{[\phi\phi]_0}(\bar{h})\big|_{\text{pert}}+(\delta h P)_{[\phi\phi]_0}(\bar{h})\big|_{\text{nonpert}}$ has no spurious poles, and the same asymptotics as $(\delta h P)_{[\phi\phi]_0}(\bar{h}) \big|_{\text{pert}}$, since 
\be
\left(\delta h_{[\phi\phi]_0}(\bar{h})\right)_{\text{nonpert}} \sim 4^{-\bar{h}}\bar{h}^{1/2-4h_{\phi}}
\ee
is exponentially damped at asymptotically large $\bar{h}$. Such exponentially damped contributions can be understood as arising from the region of integration near $\bar{z} \sim 1$, while the perturbative contributions come from expanding the integrand near $\bar{z} \sim 0$. 

We will now generalize the log matching to different external dimensions by considering the 4-point function $\<\f_1\f_2\f_2\f_1\>$. The $s$-channel OPE data is given by integrating over the $t$- and $u$-channel double discontinuities, so the anomalous dimensions are given by
\begin{equation}
\left(\delta h P\right)_{[\phi_1\phi_2]_0}(\hb)= f_{11\O}f_{22\O} \left(\delta h P\right)_{1221}(\hb)+ f_{12\O}^{2}\left(\delta h P\right)_{1212}(\hb)\;,
\end{equation}
where
\be 
{}&\left(\delta h  P\right)_{1234}(\hb)\bigg|_{\text{pert}}=  -\sin (\pi  (h_{1}+h_{4}-h_{\O})) \sin (\pi  (h_{2}+h_{3}-h_{\O})) \\ 
&\;\times \frac{\Gamma (2 h_{\O}) \Gamma(h_{\O}-h_2-h_3+1) \Gamma(h_{\O}-h_1-h_4+1)}{\pi^2 \Gamma (h_{\O}+h_{2}-h_{3}) \Gamma (h_{\O}-h_{2}+h_{3})}\\
&\;\times \frac{ \Gamma (\bar{h}+h_{1}-h_{2}) \Gamma(\bar{h}+h_3-h_4) \Gamma(\bar{h}-h_\O+h_2+h_4-1) }{\Gamma (2 \bar{h}-1) \Gamma(\bar{h}+h_{\O}-h_2-h_4+1)} \\
&\;\times \pFq{4}{3}{h_{\O}-h_{2}+h_{3},h_{\O}-h_{2}+h_{3}-\frac{d-2}{2},h_{\O}-h_{2}-h_{3}+1,h_{\O}-h_{1}-h_{4}+1}{-\bar{h}+h_{\O}-h_{2}-h_{4}+2,\bar{h}+h_{\O}-h_{2}-h_{4}+1,2 h_{\O}-\frac{d-2}{2}}{1}\;,
\ee 
\be 
\left(\delta h P\right)_{1234}(\hb)&\bigg|_{\text{nonpert}}=- \sin (\pi  (h_{1}+h_{4}-h_{\O})) \sin (\pi  (h_{2}+h_{3}-h_{\O}))\\
&\times \frac{\Gamma (2 h_{\O}) \Gamma \left(2 h_{\O}-\frac{d-2}{2}\right) }{\pi^2  \Gamma (h_{\O}+h_{2}-h_{3}) \Gamma (h_{\O}-h_{2}+h_{3})^2 \Gamma \left(h_{\O}-h_{2}+h_{3}-\frac{d-2}{2}\right)}
\\
&\times \frac{\Gamma (\bar{h}+h_{1}-h_{2}) \Gamma (\bar{h}-h_{1}+h_{2})  \Gamma (\bar{h}+h_{3}-h_{4}) \Gamma (\bar{h}-h_{3}+h_{4}) }{ \Gamma(2\bar{h}) \Gamma (2 \bar{h}-1) }
\\
&\times \frac{\Gamma(-\bar{h}+h_{\O}-h_{2}-h_{4}+1)\Gamma (\bar{h}+h_{1}+h_{2}-1) \Gamma \left(\bar{h}+h_{1}+h_{2}-\frac{d}{2}\right)}{\Gamma(\bar{h}+h_{\O}+h_2+h_4-\frac{d}{2})} \\
&\times \pFq{4}{3}{\bar{h}-h_{1}+h_{2},\bar{h}-h_{3}+h_{4},\bar{h}+h_{1}+h_{2}-1,\bar{h}+h_{1}+h_{2}-\frac{d}{2}}{2 \bar{h},\bar{h}-h_{\O}+h_{2}+h_{4},\bar{h}+h_{\O}+h_{2}+h_{4}-\frac{d}{2}}{1}\;.
\ee 

To derive these expressions from the inversion formula, we had to set $h_{1}+h_{2}=h_{3}+h_{4}$ so the $u$- and $t$-channel blocks have $\log(z)$ terms, but we left this equality implicit in the above expression.

Corrections to OPE coefficients can be derived in a similar way, by matching regular terms in $t$-channel conformal blocks. Somewhat cumbersome formulas for such corrections in general dimension and for general spin exchange were given in~\cite{Cardona:2018qrt}. In the next section we will describe an alternate and perhaps simpler approach to obtaining anomalous dimensions and OPE coefficient corrections in 3d CFTs, via dimensional reduction.

\subsection{Application to 3d CFTs}
\subsubsection{Ising CFT}
To demonstrate why nonperturbative corrections can be important, we would like to see how they affect analytic predictions for the 3d Ising CFT. We will restrict ourselves to the 4-point function $\<\s\s\s\s\>$ and extract predictions for the $[\s\s]_{0}$ scaling dimensions and OPE coefficients. We will improve the results found in \cite{Alday:2015ewa,Simmons-Duffin:2016wlq}.

For the Ising CFT we will focus on the effects of three operators, the identity operator $\mathbb{1}$, the lightest parity-even scalar $\epsilon$, and the stress-tensor $T^{\mu\nu}$. We will also use the following results from the numerical bootstrap \cite{ElShowk:2012ht,Simmons-Duffin:2016wlq} as inputs:
\be
h_{\s}{}&={}0.25907445(50)\;, \qquad
h_{\epsilon}{}={}0.7063125(50)\;, \qquad
h_{T}{}={}0.5\;,
\\
&\hspace{.25in}f_{\s\s\epsilon}{}={}1.0518537(41)\;, \qquad
f_{\s\s T}{}={}0.32613776(45)\;.
\ee

To use the inversion formula, we will use dimensional reduction to write the 3d blocks as sums of 2d blocks \cite{Hogervorst:2016hal}. Specifically, we use the expansion\footnote{Our parametrization differs from \cite{Hogervorst:2016hal}, so $\mathcal{A}^{\text{here}}_{n,j}=\mathcal{A}^{\text{there}}_{\frac{n+j}{2},\bar{h}-h+j-n}$, and our normalization is such that $c^{(d)}_{\ell}=\frac{(d-2)_\ell}{\left(\frac{d-2}{2}\right)_\ell}$ in Eq. (2.35) of \cite{Hogervorst:2016hal}.}
\bea
g^{r,s,(3d)}_{h,\bar{h}}(z,\bar{z})&=\sum\limits_{n=0}^{\infty} \ \sum\limits_{\begin{tiny}j=\max(-n,n-\ell)\end{tiny}}^{n}\mathcal{A}^{r,s}_{n,j}(h,\bar{h})g^{r,s,(2d)}_{h+n,\bar{h}+j}(z,\bar{z})\;,
\\
g^{r,s,(2d)}_{h,\bar{h}}(z,\bar{z})&=\frac{1}{1+\delta_{\bar{h}-h,0}}\left(z^h {}_{2}F_{1}(h+r,h+s,2h,z)\bar{z}^{\bar{h}} {}_{2}F_{1}(\bar{h}+r,\bar{h}+s,2\bar{h},\bar{z})+(z\leftrightarrow\bar{z})\right)\;.
\eea
In \cite{Hogervorst:2016hal} this expansion was derived in closed form for $r=s=0$, which will be sufficient for our calculations.

Since each 2d block is a sum of hypergeometrics, we can use the same techniques as when inverting a single scalar block in the previous section. Specifically, after setting $r=s=0$ and extracting the leading $z \rightarrow 0$ behavior of the hypergeometrics in the $t$-channel, we have contributions from $\SL(2,\mathbb{R})$ blocks of the form
\be
g^{(2d)}_{h_\O,\bar{h}_\O}(1-z,1-\bar{z})\bigg|_{z\rightarrow 0} =&\; \frac{1}{1+\delta_{\bar{h}_\O-h_\O,0}} \left[ - \frac{\Gamma(2 h_\O)}{\Gamma(h_\O)^2} (\log(z) + 2 \psi^{(0)}(h_\O) + 2 \gamma) + \ldots \right] \\
&\times (1-z)^{h_{\O}} k_{2\bar{h}_{\O}}(1-\bar{z}) + (h_{\O} \leftrightarrow \bar{h}_{\O}).
\ee
Here $\psi^{(0)}(z) = \Gamma'(z)/\Gamma(z)$ is the digamma function and $\gamma$ is the Euler constant. We then we find the following corrections to the OPE coefficients and anomalous dimensions after inverting the $t$-channel block: 
\newpage
\bea 
\left(\delta h P \right)^{h_\O,\bar{h}_\O}_{[\s\s]_{0}} (\bar{h})&\bigg|_{\text{pert}}=- f_{\s\s\O}^{2} \sin ^2(\pi  (2 h_{\s}-\bar{h}_{\O}))   \nn\\
&\times \frac{\Gamma(2 h_{\O}) \Gamma (\bar{h}_{\O}-2 h_{\s}+1)^2}{\pi^2 \Gamma (h_{\O})^2 } \frac{\Gamma (\bar{h})^2}{\Gamma(2\bar{h}-1) } \frac{   \Gamma(\bar{h}-\bar{h}_{\O}+2 h_{\s} -1)}{\Gamma(\bar{h}+\bar{h}_{\O}-2h_\s+1)} \nn\\
&\times \pFq{4}{3}{\bar{h}_{\O},\bar{h}_{\O},\bar{h}_{\O}-2 h_{\s}+1,\bar{h}_{\O}-2 h_{\s}+1}{2 \bar{h}_{\O},-\bar{h}+\bar{h}_{\O}-2 h_{\s}+2,\bar{h}+\bar{h}_{\O}-2 h_{\s}+1}{1}\;,
\\ \nn
\\
\left(\delta h P\right)^{h_\O,\bar{h}_\O}_{[\s\s]_{0}}(\bar{h})&\bigg|_{\text{nonpert}}=- f_{\s\s\O}^{2} \sin ^2(\pi  (2 h_{\s}-\bar{h}_{\O})) \nn\\
&\times \frac{\Gamma(2 h_{\O})\Gamma(2 \bar{h}_{\O})}{\pi^2 \Gamma (h_{\O})^2 \Gamma (\bar{h}_{\O})^2 }  \frac{ \Gamma (\bar{h})^4  }{ \Gamma(2\bar{h}-1) \Gamma(2\bar{h}) } \frac{\Gamma(-\bar{h}+\bar{h}_\O-2h_\s+1) \Gamma (\bar{h}+2 h_{\s}-1)^2}{\Gamma (\bar{h}+\bar{h}_{\O}+2 h_{\s} -1) } \nn\\
& \x\pFq{4}{3}{\bar{h},\bar{h},\bar{h}+2 h_{\s}-1,\bar{h}+2 h_{\s}-1}{2 \bar{h},\bar{h}-\bar{h}_{\O}+2 h_{\s},\bar{h}+\bar{h}_{\O}+2 h_{\s}-1}{1}\;,
\eea 
with the net contribution from a given 2d block in both the $t-$ and $u$-channels given by
\be
(\delta h P)_{[\s\s]_{0}} =&\; \frac{1+(-1)^{\bar{h}-h}}{1+\delta_{\bar{h}_\O-h_\O,0}} \left[\left(\delta h P \right)^{h_\O,\bar{h}_\O}_{[\s\s]_{0}} (\bar{h})\bigg|_{\text{pert}} + \left(\delta h P \right)^{h_\O,\bar{h}_\O}_{[\s\s]_{0}} (\bar{h})\bigg|_{\text{nonpert}} \right] \\
& + (h_\O \leftrightarrow \bar{h}_\O)\;.
\ee

\begin{figure}
  \centering
    \includegraphics[width=\textwidth]{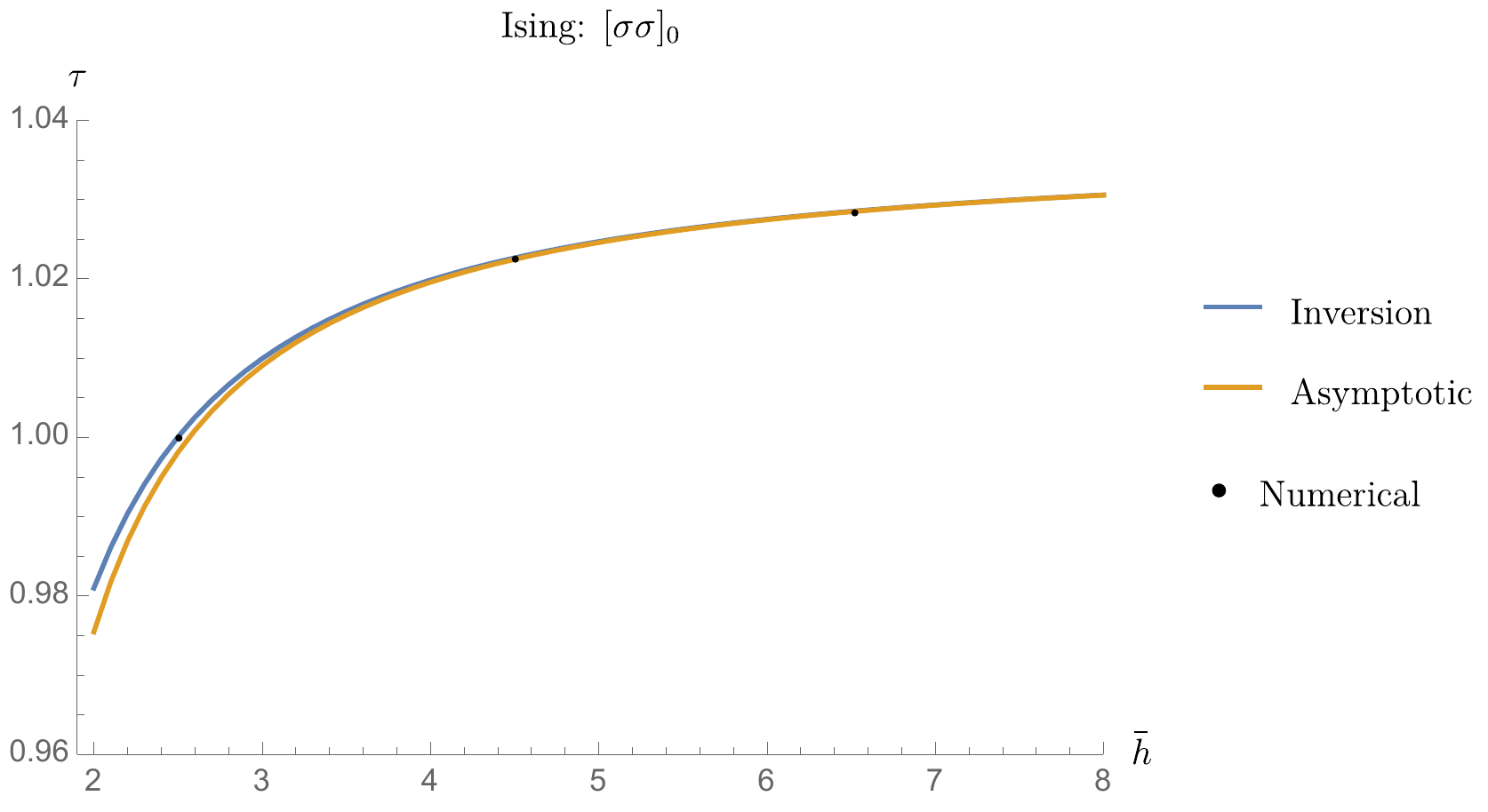}
      \caption{Spectrum for $[\sigma\sigma]_{0}$ in the Ising CFT derived using the inversion formula, asymptotic lightcone expansion, and numerical bootstrap. Numerical data is taken from \cite{Simmons-Duffin:2016wlq}. The curves in this and later plots are obtained by matching at $z=.1$.}
\end{figure}

Similarly, by matching regular terms we obtain the corrections to the OPE coefficients
\be
\delta P_{[\s\s]_{0}} =&\;  \frac{1+(-1)^{\bar{h}-h}}{1+\delta_{\bar{h}_\O-h_\O,0}} \left(2 \psi^{(0)}(h_\O) + 2 \gamma\right)  \left[ \left(\delta h P \right)^{h_\O,\bar{h}_\O}_{[\s\s]_{0}} (\bar{h})\bigg|_{\text{pert}} + \left(\delta h P \right)^{h_\O,\bar{h}_\O}_{[\s\s]_{0}} (\bar{h})\bigg|_{\text{nonpert}} \right]\\&+ (h_\O \leftrightarrow \bar{h}_\O)\;.
\ee

Note that if we take $\O = \mathbb{1}$ to be the identity operator and take the limit \mbox{$h_\O = \bar{h}_\O \rightarrow 0$} (as well as set $f_{\s\s\mathbb{1}} = 1$), then we reproduce the expected identity contribution $P_{[\s\s]_{0}} = (1+ (-1)^{\bar{h}-h}) S_{-2h_\s}(\bar{h})$.

At finite spin and finite anomalous dimensions one does not expect that it is sufficient to match the terms logarithmic and regular in $z$ to obtain the precise OPE data. Although inverting individual operators produces factors of $z^{h_{[\s\s]_{n,\ell}}}$ and $z^{h_{[\s\s]_{n,\ell}}}\log z$, we know the exact generating function $C^t(z,\bar{h})$ at small $z$ is~\cite{Caron-Huot:2017vep}:
\begin{align}
C^t(z,\bar{h})=C_{[\s\s]_{0}}(\bar{h})z^{2h_{\s}+\delta h_{[\s\s]_{0}}(\bar{h})}+... \text{ ,}
\end{align}
where we ignore terms subleading in $z$.

We can then extract the anomalous dimension via:
\begin{align}
\delta h_{[\s\s]_{0}}(\bar{h})= \lim\limits_{z\rightarrow 0}\frac{(z\partial_{z}-2h_{\s})C^t(z,\bar{h})}{C^t(z,\bar{h})}\;,
\end{align}
which we in practice evaluate by evaluating the generating function at small but finite $z$. We find the OPE coefficients in a similar way by taking our value for $\delta h_{[\s\s]_{0}}(\bar{h})$ and using:
\begin{align}
C_{[\s\s]_{0}}(\bar{h})=\lim\limits_{z\rightarrow 0}\frac{C^t(z,\bar{h})}{z^{2h_{\s}+\delta h_{[\s\s]_{0}}(\bar{h})}}\;,
\end{align}
where we once again evaluate the right-hand side at small but finite $z$.

In evaluating these expressions one wishes to take $z$ small, but not too small so as to avoid neglected terms with higher powers of $\log z$ from becoming important. In~\cite{Simmons-Duffin:2016wlq} it was found that $z=.1$ is a good choice for the Ising model (there called $\bar{y}_0$), so we will present results at this value in our initial analysis. In future work it may be helpful to further optimize the matching value of $z$. As more operators are included one should also see that the results become less and less sensitive to this choice.

Now the procedure should be clear: we can expand the 3d blocks as sums of 2d blocks and invert each block term by term. This procedure is sufficient to extract finite-spin data from the Lorentzian inversion formula. In practice we find that we need to expand to at most 10 to 15 orders in the 2d expansion such that the errors introduced by truncating this expansion are smaller than the errors from the numerical input. 

With this data and the above expressions, we can extract $P_{[\s\s]_{0}}$ and $\delta h_{[\s\s]_{0}}$, but we have to do a little more work to extract the physical OPE coefficients and scaling dimensions. To find the scaling dimensions, we need to solve the equation
\be
\bar {h}-2h_{\s}-\delta h_{[\s\s]_{0}}(\bar{h})=\ell\;,
\ee
where $\ell$ is the spin of the local, double-twist operator. As the anomalous dimensions are expressed in terms of ${}_{4}F_{3}$ hypergeometric functions, we will solve this equation numerically. We then calculate the physical OPE coefficients $f_{\s\s[\s\s]_{0}}$ using the relation
\be
f^{2}_{\s\s[\s\s]_{0}}\approx\left(1-\frac{\partial \delta h_{[\s\s]_{0}}(\bar{h})}{\partial\bar{h}}\right)^{-1}P_{\s\s[\s\s]_{0}}\;,
\ee
where the Jacobian appears because we need to take residues of the OPE function $c(\Delta,\ell)$ in terms of $\Delta$ at fixed spin $\ell$.
	
\begin{figure}
  \centering
    \includegraphics[width=\textwidth]{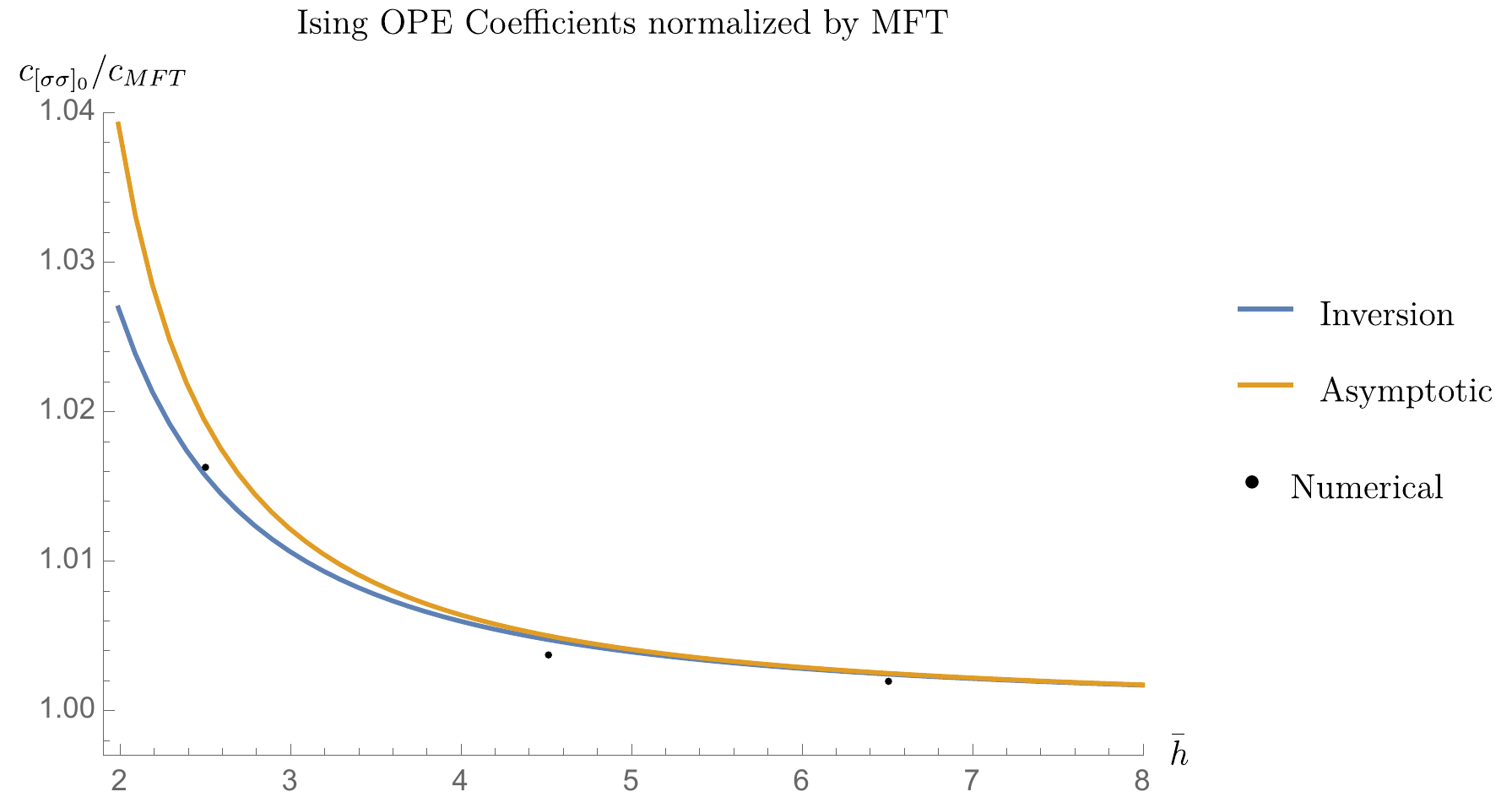}
      \caption{OPE coefficients  $f_{\s\s[\s\s]_{0}}$ in the Ising CFT. Numerical data is taken from \cite{Simmons-Duffin:2016wlq} and the OPE coefficients are normalized by dividing by the mean field theory OPE coefficients.}
\end{figure}

A comparison of results from the numerical bootstrap~\cite{Simmons-Duffin:2016wlq}, the leading asymptotic lightcone bootstrap (\ref{eq:asymptoticP},~\ref{eq:asymptoticdh}),\footnote{We do not go to higher orders in the lightcone bootstrap because the expansion then introduces spurious poles at low spin.} and the inversion formula result, can be found in Table \ref{table:IsingInversion}. We focus here on how accurately we can reproduce the low-spin data. We see that in all cases, including the nonperturbative effects from the inversion formula leads to more accurate results. This is clearest for the scaling dimensions, where we have at least an extra digit of precision for the lightest spin-$2$ and spin-$4$ operators. 

This improvement is especially marked for the stress-tensor and gives additional evidence that the stress-tensor should be thought of as a double-twist operator composed of two $\sigma$ operators. We see a similar improvement for the OPE coefficients, although it is smaller in comparison to the dimensions. The errors listed come from the errors in the numerical input and do not include errors from truncating the operator product expansion to include only a few light operators. As we include more operators beyond $\epsilon$ and $T^{\mu\nu}$ we expect the results to improve even further, but we postpone this analysis to a future study where effects related to operator mixing can also be taken into account.

\begin{table}

   \begin{tabularx}{\textwidth}{XXXXXX}
	\hline\hline
	{} & \textbf{Numerics}&  \textbf{Inversion${}_{z=.1}$}& \textbf{Inversion${}_{\log z}$}& \textbf{Lightcone${}_{z=.1}$}  & \textbf{Lightcone${}_{\log z}$}\\\hline
	$\tau_{[\s\s]_{0,2}}$& 1 & 1.000060(2) & 0.998459(4) &0.998082(4) &0.9962944(46) \\ 
	$\tau_{[\s\s]_{0,4}}$&1.022665(28)&1.0226890(7)&1.022472(3)&1.022510(3)&1.0222880(28)\\ 
	$f_{\s\s[\s\s]_{0,2}}$&0.32613776(45) &0.325981(1) &0.3262377(9)&0.327398(1) &0.3277057(10)\\ 
	$f_{\s\s[\s\s]_{0,4}}$&0.069076(43)&0.0691405(2)&0.0691445(2) &0.0691630(2)&0.0691671(2)\\ \hline \hline
\end{tabularx}
\caption{\label{table:IsingInversion}We list results for the twists and OPE coefficients for the double-twist family $[\s\s]_{0,\ell}$ in the 3d Ising model by either matching at $z=.1$ or using the na\"ive $\log z$ matching valid for perturbative anomalous dimensions.  Approximate errors come from numerical input.}
	\end{table}\

%\begin{table}
%
%   \begin{tabularx}{\textwidth}{XXXX}
%	\hline\hline
%	{} & \textbf{Numerics}&  \textbf{Inversion}& \textbf{Lightcone}\\\hline
%	$\tau_{[\s\s]_{0,2}}$& 1 & 0.998459(4) &0.9962944(46) \\ 
%	$\tau_{[\s\s]_{0,4}}$&1.022665(28)&1.022472(3)&1.0222880(28)\\ 
%	$f_{\s\s[\s\s]_{0,2}}$&0.32613776(45) &0.3262377(9) &0.3277057(10)\\ 
%	$f_{\s\s[\s\s]_{0,4}}$&0.069076(43)&$0.0691445(2)$&0.0691671(2)\\ \hline \hline
%\end{tabularx}
%\caption{\label{table:IsingInversionV1}We list results for the twists and OPE coefficients for the double-twist family $[\s\s]_{0,\ell}$ in the 3d Ising model from $\log$ and power law matching. Approximate errors come from numerical input.}
%	\end{table}\
%
%\begin{table}
%
%   \begin{tabularx}{\textwidth}{XXXX}
%	\hline\hline
%	{} & \textbf{Numerics}& \textbf{Inversion}& \textbf{Lightcone}\\\hline
%	$\tau_{[\s\s]_{0,2}}$& 1 & 1.000060(2) &0.998082(4) \\ 
%	$\tau_{[\s\s]_{0,4}}$&1.022665(28)&1.0226890(7)&1.022510(3)\\ 
%	$f_{\s\s[\s\s]_{0,2}}$&0.32613776(45) &0.325981(1) &0.327398(1)\\ 
%	$f_{\s\s[\s\s]_{0,4}}$&0.069076(43)&$0.0691405(2)$&0.0691630(2)\\ \hline \hline
%\end{tabularx}
%\caption{\label{table:IsingInversionV2}Results for the 3d Ising model using the (truncated) exact sum rule evaluated at $z=.1$.}
%	\end{table}\	

\subsubsection{O(2) model}

We can repeat the above analysis, but now for the O(2) vector model. We will study the 4-point function of fundamental scalars $\<\f^{i}\f^{j}\f^{k}\f^{\ell}\>$, which we can decompose in terms of the exchanged global symmetry representations as
\begin{multline}
x_{12}^{2\Delta_{\f}}x_{34}^{2\Delta_{\f}}\<\f_{i}\f_{j}\f_{k}\f_{l}\>=\delta_{ij}\delta_{kl}I(u,v)+(\delta_{il}\delta_{jk}-\delta_{ik}\delta_{jl})A(u,v)
\\ + \left(\delta_{il}\delta_{jk}+\delta_{ik}\delta_{jl}-\delta_{ij}\delta_{kl}\right)S(u,v)\;,
\end{multline}
where $I$, $A$, and $S$ correspond to contributions from exchanged operators that transform in the singlet, antisymmetric, and symmetric traceless representation of O(2), respectively. 

If we collect them into a vector $\vec{Z}(u,v)=\{I(u,v),A(u,v),S(u,v)\}$, then $(1,i)\leftrightarrow (3,k)$ crossing implies
\begin{subequations}
\be 
\left(\frac{u}{v}\right)^{\Delta_{\f}}\vec{Z}(u,v)=M\cdot \vec{Z}(u,v)\;,
\ee 
for
\be 
M=\left( \begin{array}{ccc}
	\frac{1}{2} & \frac{1}{2} &  1 \vspace{.085cm} \\
	\frac{1}{2} & \frac{1}{2} & -1  \vspace{.085cm} \\
	\frac{1}{2} & -\frac{1}{2} & 0 
\end{array}
\right). \label{eqn:matO2}
\ee 
\end{subequations}

We will use the following results from the numerical bootstrap \cite{Kos:2013tga,Kos:2015mba,Kos:2016ysd}:
\begin{subequations}
\begin{alignat}{2}
h_{\f}&=0.25963(16)\;,
& h_{\f^{2}}&=0.7559(13)\;,
\\
h_{t}&=0.6179(16)\;, 
& f_{\f\f\f^{2}}&=0.68726(65)\;,
\\
f_{\f\f J}&=0.52558(46)\;,\quad  &
f_{\f\f T}&=0.23146(16)\;.
\end{alignat}
\end{subequations}

Here $t$ refers to the lightest symmetric, traceless scalar in the $\f_{i} \times \f_{j}$ OPE. There is one crucial piece of OPE data missing, the OPE coefficient $f_{\f\f t}$, although there are estimates from the $\epsilon$-expansion \cite{Dey:2016mcs}, which yield
\begin{align}
f_{\f\f t} &\approx \{0.8944,\, 0.8246,\, 0.8850\} \quad \text{at} \quad \{\O(\epsilon),\, \O(\epsilon^2),\, \O(\epsilon^3)\}.\label{eq:O2opeRange}
\end{align}

Using this data as input, we can calculate the low-spin spectrum for the O(2) vector model using either the asymptotic lightcone bootstrap or the inversion formula. In our calculations we expanded to $12^{th}$ and $20^{th}$ order in the 2d conformal blocks to obtain converged results for the stress-tensor and conserved current OPE data, respectively. The results are shown in Table \ref{table:O2Inversion}.

\begin{figure}
  \centering
    \includegraphics[width=\textwidth]{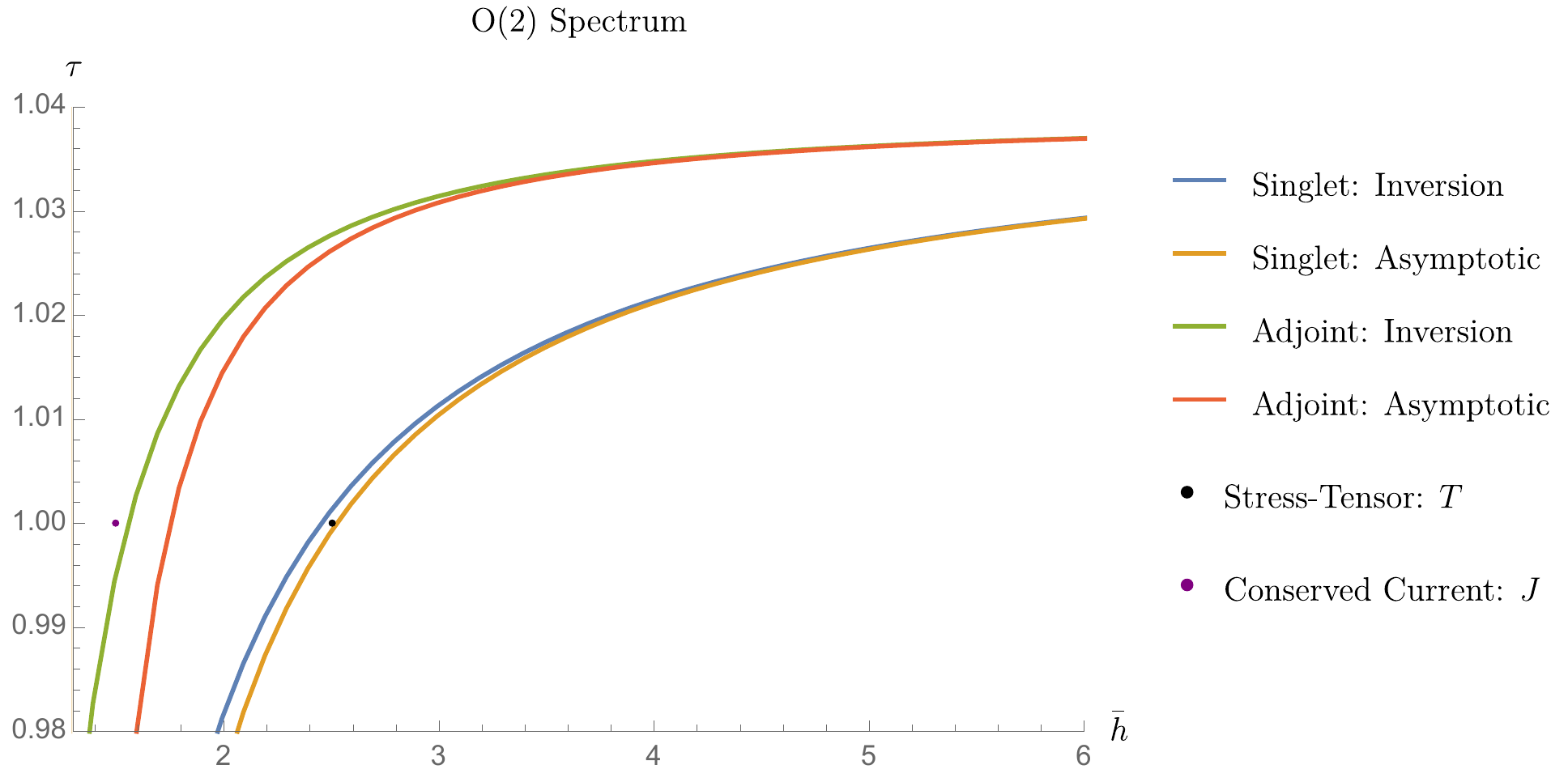}
      \caption{Spectrum of $[\f\f]_{0}^{(I)}$ and $[\f\f]_{0}^{(A)}$ in the O(2) model. The black dots corresponds to the stress tensor and conserved current which have twist one.}
\end{figure}

We see that the inversion formula in general gives more accurate results for both the conserved current $J^{\mu}$ and the stress-tensor $T^{\mu\nu}$. The improvement is particularly large for $c_{\f\f[\f\f]^{A}_{0,1}}$, or the coupling between two scalars and $J^{\mu}$. One reason the inversion formula gives an improved estimate for this OPE coefficient is because it also gives a much more accurate result for $\tau_{[\f\f]^{(A)}_{0,1}}$, which is used as input in the calculation of the OPE coefficient. Finally, we should note that the inversion formula is only guaranteed to hold for spin $J>1$, but we see for the O(2) vector model it likely holds down to at least $J=1$.\footnote{ There are also indications that analyticity holds all the way down to $J=0$ in the Ising CFT \cite{SCH:WIP}.} 

The one exception appears to be the twist of the stress-tensor itself, for which the lightcone analysis gives a result which is slightly closer to the exact answer. We expect this is an artifact of truncating the $t$-channel expansion: as we include more operators the results for the twist will decrease which will push the lightcone result further from the exact result.

\begin{table}
   \begin{tabularx}{\textwidth}{XXXXXX}
	\hline\hline
	{} & \textbf{Numerics}& \textbf{Inversion${}_{z=.1}$} & \textbf{Inversion${}_{\log z}$} & \textbf{Lightcone${}_{z=.1}$} & \textbf{Lightcone${}_{\log z}$}\\
	\hline
	$\tau_{[\f\f]^{(I)}_{0,2}}$  &  1&1.0012(24) &0.9996(26) & 0.9992(25) & 0.9973(27)\\ 
	$\tau_{[\f\f]^{(A)}_{0,1}}$ & 1 & 0.9958(60)& 0.9933(66) & 0.9480(90)& 0.933(13) \\ 
	$f_{\f\f[\f\f]^{(I)}_{0,2}}$  &0.231462(16) &0.23128(46)&0.23147(48)&0.23231(50)&0.23254(52)\\
	$f_{\f\f[\f\f]^{(A)}_{0,1}}$  & 0.52558(46) & 0.5270(32)& 0.5286(36)&0.6005(97) &0.630(17)\\ \hline\hline 
\end{tabularx}
\caption{\label{table:O2Inversion}We list results for the twists and OPE coefficients for the double-twist family $[\f\f]_{0,\ell}$ in the 3d O(2) model by either matching at $z=.1$ or using the na\"ive $\log z$ matching valid for perturbative anomalous dimensions. The errors are approximate and come from both numerical input and from using the lower and upper values in (\ref{eq:O2opeRange}).}
	\end{table}

%\begin{table}
%   \begin{tabularx}{\textwidth}{XXXX}
%	\hline\hline
%	{} & \textbf{Numerics}& \textbf{Inversion} & \textbf{Lightcone}\\
%	\hline
%	$\tau_{[\f\f]^{(I)}_{0,2}}$  &  1&0.9996(26)& 0.9973(27)\\ 
%	$\tau_{[\f\f]^{(A)}_{0,1}}$ & 1 & 0.9933(66) & 0.933(13) \\ 
%	$f_{\f\f[\f\f]^{(I)}_{0,2}}$  &0.231462(16) &0.23147(48)&0.23254(52)\\
%	$f_{\f\f[\f\f]^{(A)}_{0,1}}$  & 0.52558(46) & 0.5286(36)&0.630(17)\\ \hline\hline 
%\end{tabularx}
%\caption{\label{table:O2InversionV1}We list results for the twists and OPE coefficients for the double-twist family $[\f\f]_{0,\ell}$ in the 3d O(2) model from $\log$ and power law matching. The errors are approximate and come from both numerical input and from using the lower and upper values in (\ref{eq:O2opeRange}).}
%	\end{table}
%	
%\begin{table}
%   \begin{tabularx}{\textwidth}{XXXX}
%	\hline\hline
%	{} & \textbf{Numerics}& \textbf{Inversion} & \textbf{Lightcone}\\
%	\hline
%	$\tau_{[\f\f]^{(I)}_{0,2}}$  &  1&1.0012(24)& 0.9992(25)\\ 
%	$\tau_{[\f\f]^{(A)}_{0,1}}$ & 1 & 0.9958(60) & 0.9480(90) \\ 
%	$f_{\f\f[\f\f]^{(I)}_{0,2}}$  &0.231462(16) &0.23128(46)&0.23231(50)\\
%	$f_{\f\f[\f\f]^{(A)}_{0,1}}$  & 0.52558(46) & 0.5270(32)&0.6005(97)\\ \hline\hline 
%\end{tabularx}
%\caption{\label{table:O2InversionV2}Same as above except using the (truncated) exact sum rule evaluated at $z=.1$.}
%	\end{table}

We can also take a different point of view and use the inversion formula to make a prediction for $f_{\f\f t}$. For example, if we require that the inversion formula reproduces the exact twist of $T^{\mu\nu}$ then we find the following range:
\be
\qquad f_{\f\f t}\in(0.857,0.951)\,,
\ee
with a central value of approximately $f_{\f\f t}=0.9038$. Using results from Monte Carlo~\cite{Campostrini:2006ms} as input, setting $h_{\f}=0.259525(50)$, $h_{\f^{2}}=0.75562(11)$, and $h_{t}=0.6180(5),$\footnote{This range comes from comparing the bootstrap data in Figure 9 of~\cite{Kos:2015mba} with the Monte Carlo allowed region.} and repeating the above analysis, the window shrinks to
\be
\qquad f_{\f\f t}\in (0.883,0.901)\,.
\ee

\begin{figure}
  \centering
    \includegraphics[width=\textwidth]{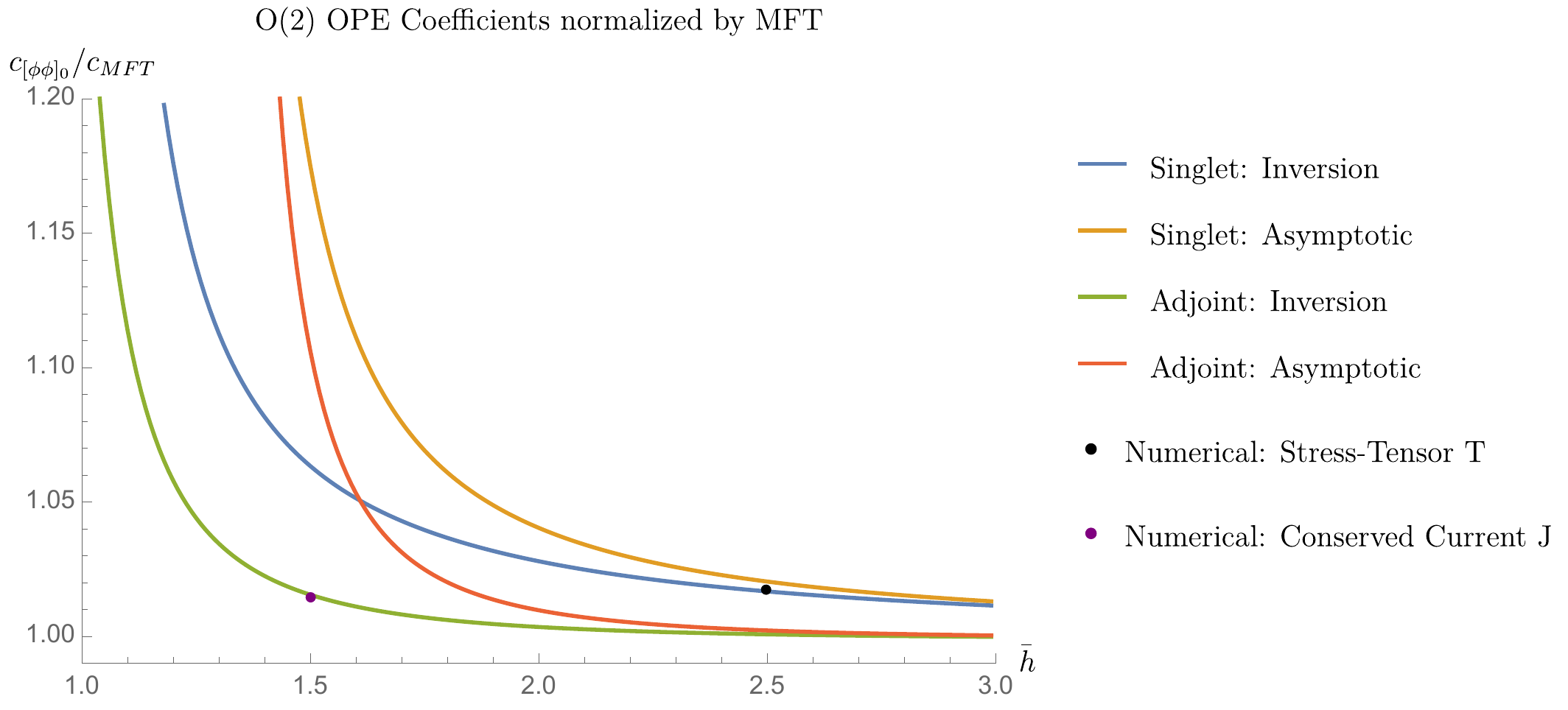}
      \caption{OPE coefficients for $f_{\f\f[\f\f]^{(I)}_{0}}$ and $f_{\f\f[\f\f]^{(A)}_{0}}$ in the O(2) model. Numerical data is taken from \cite{Kos:2013tga} and the OPE coefficients are normalized by dividing by the mean field theory OPE coefficients.}
\end{figure}

By including more operators in the inversion formula or the effects of operator mixing it should be possible to improve the above results further. It will be interesting to understand which operators need to be included in order to reproduce the current beyond the $10^{-3}$ level. It would also be interesting to extend this work to higher orders in the small-$z$ expansion to understand the higher-twist families. We plan to return to this in future work.

\section{Fermion Analytic Bootstrap}
\label{sec:fermion}

In the remainder of the paper we will move to 4-point functions of fermions $\<\psi\psi\psi\psi\>$ in 3d. Our focus for now will be to compute the first several perturbative $1/\bar{h}$ corrections to the $[\psi\psi]_0$ coefficients and anomalous dimensions, generalizing the method of section~\ref{sec:LightconeBootstrapReview} to fermions. As illustrated in~\cite{Simmons-Duffin:2016wlq}, such perturbative calculations are in many cases numerically sufficient and also quite useful in providing consistency checks on formulas obtained from the inversion formula approach. In future work it will also be interesting to generalize the inversion formula and nonperturbative resummations studied in section~\ref{sec:inversion} to fermion correlators.

\subsection{Fermion conformal blocks and the crossing equation}
We will be using the method of \cite{Iliesiu:2015qra} to generate the fermion conformal blocks by hitting the four-scalar conformal block with differential operators in the embedding space.\footnote{A review of the embedding space formalism can be found in Appendix \ref{sec:EmbeddingReview}.} Specifically, a contribution to $\<\psi\psi\psi\psi\>$ takes the form:
\\
\begin{multline}
\left(\frac{X_{24}}{X_{14}}\right)^{\frac{\Delta_{12}}{2}}\left(\frac{X_{14}}{X_{13}}\right)^{\frac{\Delta_{34}}{2}}\frac{t_Ig^{I;a,b}_{h,\bar{h}}(z,\bar{z})}{X_{12}^{\frac{\Delta_1+\Delta_2+1}{2}}X_{34}^{\frac{\Delta_3+\Delta_4+1}{2}}}\\=\DD_a\widetilde{\DD}_b\left[
\left(\frac{X_{24}}{X_{14}}\right)^{\frac{\Delta_{12}}{2}}\left(\frac{X_{14}}{X_{13}}\right)^{\frac{\Delta_{34}}{2}}\frac{g_{h,\bar{h}}(z,\bar{z})}{X_{12}^{\frac{\Delta_1+\Delta_2}{2}}X_{34}^{\frac{\Delta_3+\Delta_4}{2}}}
\right]\;,\label{eq:crossingStandardForm}
\end{multline}
where $t_I$ are different 4-point structures in embedding space, $a,b$ are indices denoting possible 3-point structures, and we have suppressed the dependence of the blocks on the external dimensions. The operators $\DD_a$ are defined as follows:
\bea[eq:DifferentialOperators]
&\DD_1=\<S_1S_2\>\shift{1^+2^+}\;,\\
&\DD_2=-\<S_1\frac{\delta}{\delta X_1}\frac{\delta}{\delta X_2}S_2\>\shift{1^-2^-}\;,\\
&\DD_3=\<S_1\frac{\delta}{\delta X_1}S_2\>\shift{1^-2^+}-\<S_2\frac{\delta}{\delta X_2}S_1\>\shift{1^+2^-}\;,\\
&\DD_4=\<S_1\frac{\delta}{\delta X_1}S_2\>\shift{1^-2^+}+\<S_2\frac{\delta}{\delta X_2}S_1\>\shift{1^+2^-}\;,
\eea
where $\shift{a^i b^j}$ are operators which shift external dimensions such that
\be 
\shift{a^i b^j}: \{\Delta_a,\Delta_b\}\rightarrow \{\Delta_a+\frac{i}{2},\Delta_b+\frac{j}{2}\}\;.
\ee 
The corresponding operators acting on the points $(X_3,X_4)$ are found via the replacement
\be 
\widetilde{\DD}_a\coloneqq\DD_a\evaluated_{(1,2)\rightarrow(3,4)}\;.
\ee

An alternative basis denoted by $\mathcal{D}_{i}$ was used in \cite{Iliesiu:2015qra}, given by:\footnote{There is also an additional term in $\D_4$ which vanishes for identical external operators.}
\bea[changebasis]
\D_1=&\;\DD_1\;,\\
\D_2=&\;\frac{1}{4(\hb-h)(\hb+h-1)}\left(\DD_2-(2h+2\Delta_{\psi}-4)(2h-2\Delta_{\psi}+1)\DD_1\right)\;,\\
\D_3=&\;\frac{1}{2(\hb+h-1)}\DD_3\;,\\
\D_4=&\;\frac{1}{2(\hb-h)}\DD_4\;.
\eea

These two bases have different merits. One nice feature of the $\DD_i$ basis is that the operators are independent of $(h,\bar{h})$, so there is a clean separation between the calculation of double-twist data and the differential operators, i.e. we do not want these operators to also depend on the anomalous dimensions. By comparison $\D_i$ generates the most natural basis of embedding space, 3-point tensor structures. We find that it is most convenient to use the $\DD_i$ basis when performing the calculations and presenting the results.

The operators $\DD_1$ and $\DD_2$ generate parity-even structures, whereas $\DD_3$ and $\DD_4$ generate parity-odd ones. When the external fermions are identical, there are also selection rules on the spins of the exchanged operators: $\DD_{1}$, $\DD_{2}$, and $\DD_{3}$ are associated to operators of even spin while odd spins are associated with $\DD_4$.

Now let us write down the condition from crossing symmetry. For convenience we will define the prefactor $\mathsf{p}$ as
\be
\pre\left(\Delta_1,\Delta_2,\Delta_3,\Delta_4\right) \equiv 
\left(\frac{X_{24}}{X_{14}}\right)^{\frac{\Delta_{12}}{2}}\left(\frac{X_{14}}{X_{13}}\right)^{\frac{\Delta_{34}}{2}} X_{12}^{-\frac{\Delta_1+\Delta_2}{2}}X_{34}^{-\frac{\Delta_3+\Delta_4}{2}}\;,
\ee
and introduce the shorthand notations
\bea\pre_i &\equiv\pre\left(\Delta_1,\Delta_2,\Delta_3,\Delta_4\right)\;,\\
\pre_\psi &\equiv\pre\left(\Delta_\psi+\frac{1}{2},\Delta_\psi+\frac{1}{2},\Delta_\psi+\frac{1}{2},\Delta_\psi+\frac{1}{2}\right)\;.
\eea 
Then we can write a more compact form of the conformal block for identical fermions:
\be
t_Ig^{I;a,b}_{h,\bar{h}}(z,\bar{z})=\frac{\left(\DD_a\widetilde{\DD}_b\left[\pre_i\;
	g_{h,\bar{h}}(z,\bar{z})
	\right]\right)_{\Delta_i\rightarrow\Delta_\psi}}{\pre_\psi}\;,\label{eq:FermionConformalBlock}
\ee
and crossing symmetry implies
\be
\pre_\psi\sum\limits_\ope t_I  P _\ope^{a,b}g^{I;ab}_{h,\hb}(z,\zb)
=-\left(\pre_\psi t_I\right)\evaluated_{1\leftrightarrow 3}\sum\limits_\ope   P _\ope^{a,b}g^{I;ab}_{h,\hb}(1-\zb,1-z)\;,\label{crossing_fermion}
\ee 
where we use $P$ for the coefficients in the $\DD_{a}$ differential basis.

To simplify some expressions, we will first consider the contributions of double-twist operators in the $(12)\rightarrow (34)$ OPE. Then later we will take $1\leftrightarrow 3$ so that they appear in the $t$-channel, and match their sum to the contributions of individual $s$-channel blocks, as in section~\ref{sec:LightconeBootstrapReview}.

The 3-point structures will be unimportant in the following discussion so we will simplify the notation of the left-hand side of (\ref{crossing_fermion}) to
\be
\sum\limits_{\substack{\text{different}\\\text{ structures}}}\sum\limits_\ope\left(\DD\widetilde{\DD}\left[\pre_i P _\ope g_{h,\hb}(z,\zb)\right]\right)\evaluated_{\Delta_i\rightarrow\Delta_\psi}\;.
\ee 

We can also interchange the order of differential operators and summation over relevant operator families. Since $\pre$ is independent of the exchanged operator, the double-twist sum reduces to
\be
\sum\limits_{\substack{\text{different}\\\text{ structures}}}\left(\DD\widetilde{\DD}\left[\pre_i \sum\limits_\ope P _\ope g_{h,\hb}(z,\zb)\right]\right)_{\Delta_i\rightarrow\Delta_\psi}\;.
\ee 

To go any further, we need to specify which operators $\ope$ must appear to reproduce the lightcone limit of the crossed channel. In particular, we know that an infinite sum of double-twist operators is needed to reproduce the identity operator in the crossed channel~\cite{Simmons-Duffin:2016wlq,Komargodski:2012ek,Fitzpatrick:2012yx}. The required operators and their quantum numbers are schematically shown in Table~\ref{table:DoubleTwistFamilies}.\footnote{Their detailed form does not matter for our purposes as the bootstrap methods are only sensitive to twist accumulation points and parities. The quantum numbers can be derived in a variety of ways. E.g., the parities follow from the alternating structure of the tensor product $j_1^+\otimes j_2^+ =(j_{1}+j_2)^+\oplus (j_1+j_2-1)^-\oplus (j_1+j_2-2)^+\oplus \cdots \oplus \abs{j_1-j_2}^{+|-}$, where the $\pm$ superscripts denote the parity of the spin-$j$ representation of the Pin group. The spin selection rules follow from the relation $\psi_\a \partial_{\mu_1}\dots \partial_{\mu_l} \partial^{2n} \psi_\b = (-1)^{\ell+1}\psi_\b \partial_{\mu_1}\dots \partial_{\mu_l} \partial^{2n} \psi_\a + \text{(total derivatives)}$, obtained via integration by parts.}

\begin{table}
\begin{tabularx}{\textwidth}{ XXXX }
	\hline\hline
	\textbf{Family} & \textbf{Twist} & \textbf{Parity} & \textbf{Spin}\\[.01in]
	\hline
	$[\psi_{(\alpha}\psi_{\beta)}]_{n,l}$&$2\Delta_\psi+2n-1$&Even& $l\ge 2$, Even
	\\[.01in]	
	$[\psi_\alpha\psi^\alpha]_{n,l}$&$2\Delta_\psi+2n$&Odd
	& $l\ge 0$, Even
	\\[.01in]
	$[\psi_{(\rho}\partial^\rho_{\;\a}\psi_{\beta)}]_{n,l}$&$2\Delta_\psi+2n$&Odd &
	$l\ge 1$, Odd
	\\[.01in]
	$[\psi_\alpha(\partial\psi)^\alpha]_{n,l}$&$2\Delta_\psi+2n+1$&Even
	&
	$l\ge 0$, Even
	\\\hline\hline
\end{tabularx}
	\caption{\label{table:DoubleTwistFamilies}We list the different double-twist families for Majorana fermions in 3d, with their twist accumulation points, parities, and spins.}
\end{table}

To remove clutter, we will denote different double-twist families generically as $[\psi\psi]_n$ below. Their contribution then reads as
\be 
\sum\limits_{\substack{\text{different}\\\text{ structures}}}\Bigg(\DD\widetilde{\DD}\bigg[\pre_i\sum\limits_{n=0}^\infty\sum\limits_{l=0}^\infty P _{[\psi\psi]_n}
g_{h_{[\psi\psi]_n},\,\bar{h}_{[\psi \psi]_n}}(z,\zb)
\bigg]\Bigg)_{\Delta_i\rightarrow\Delta_\psi}\;,
\ee 
where the summation over all relevant families $[\psi\psi]_n$ appearing in Table~\ref{table:DoubleTwistFamilies} is implicit.

Restoring possible dependence on the external dimension differences $r,s$ (which arise from the shift operators $\Pi$), we will now use Eq.~(\ref{eq:SL2R}) in a slightly modified form after applying $g(z,\zb)=g(\zb,z)$ symmetry:
\be
g_{h,\hb}^{r,s}(z,\zb)=\sum\limits_{n=0}^\infty\sum\limits_{j=-n}^n A_{n,j}^{r,s}(h,\hb)\bar{z}^{h+n} k^{r,s}_{2(\bar{h}+j)}(z)\;.
\ee
By expanding
\be 
h_{[\psi\psi]_n}=h_{[\psi\psi]_0}+n+\delta h_{[\psi\psi]_n}\;,
\ee
the leading part of the double-twist sum at small $\bar{z}$ can then be rewritten as
\be
	\sum\limits_{\substack{\text{different}\\\text{ structures}}}\Bigg(\DD\widetilde{\DD}\bigg[\pre_i\sum\limits_{l=0}^\infty P_{[\psi\psi]_0}\zb^{h_{[\psi\psi]_0}+\delta h_{[\psi\psi]_0}} k^{r,s}_{2\hb_{[\psi\psi]_0}}(z)
	\bigg]\Bigg)_{\Delta_i\rightarrow\Delta_\psi}\left(1+\order{{\zb}}\right)\;.
\ee

To reproduce $\log$ terms in the crossed channel, we will expand to linear order in the anomalous dimension:\footnote{Higher order terms in $\delta h$ are matched with multi-twist operators in the $s$-channel.}
\be
\sum\limits_{\substack{\text{different}\\\text{ structures}}}\Bigg( \DD\widetilde{\DD}\bigg[\pre_i\;\zb^{h_{[\psi\psi]_0}}\bigg\{
\sum\limits_{l=0}^\infty P _{[\psi\psi]_0} k^{r,s}_{2\hb_{[\psi\psi]_0}}(z)
+\log(\zb)\sum\limits_{l=0}^\infty P _{[\psi\psi]_0}\delta h_{[\psi\psi]_0}
k^{r,s}_{2\hb_{[\psi\psi]_0}}(z)
\bigg\}\bigg]\Bigg)_{\Delta_i\rightarrow\Delta_\psi}.
\label{eq:tChannelExpression1}
\ee

In \cite{Simmons-Duffin:2016wlq} it was shown how to sum over $\tfo$ hypergeometric functions to reproduce terms Casimir-singular in $z$:
\be 
\sum_{l=0}^\infty\frac{\partial \hb}{\partial l}S_a^{r,s}(\hb)k^{r,s}_{2\hb}(1-z)
=\left(\frac{z}{1-z}\right)^{a}+[\cdots]_z\;, \label{eq:SomeOverNandL}
\ee 
which is the generalization of \equref{eq:SL2R_Sum} to non-identical external scaling dimensions. The explicit form of $S_a^{r,s}(h)$ is given in \equref{eq:ExpansionCoefficientSGeneralized}, though it will not be necessary for the following calculations.

As explained in section~\ref{sec:LightconeBootstrapReview}, we make the ansatz
\bea[eq:OPEAnsatz]
 P _{[\psi\psi]_0}^{a,b}(\bar{h})=&\left(\frac{\partial \hb_{[\psi\psi]_0}}{\partial l}\right)\sum\limits_{\{i\}}A_{a,b,i}S_i^{0,0}(\hb_{[\psi\psi]_0})\;,
 \label{eq:Definition of coefficient A}
 \\
(\delta hP )^{a,b}_{[\psi\psi]_0}(\bar{h})=&\left(\frac{\partial \hb_{[\psi\psi]_0}}{\partial l}\right)\sum\limits_{\{j\}}B_{a,b,j}S_j^{0,0}(\hb_{[\psi\psi]_0})\;,
\eea
and insert this into \equref{eq:tChannelExpression1} to obtain
\be
	\frac{1}{2}\sum\limits_{\substack{\text{different}\\\text{ structures}}}\left( \DD\widetilde{\DD}\left[\pre_i\zb^{h_{[\psi\psi]_0}}\left(
	\sum\limits_{\{i\}}A_{a,b,i}\left(\frac{1-z}{z}\right)^{i}
	+\log(\zb) 
	\sum\limits_{\{j\}}B_{a,b,j}\left(\frac{1-z}{z}\right)^{j}
	\right)\right]\right)_{\Delta_i\rightarrow\Delta_\psi}\;. \label{eq:CrossedChannelFinalForm}
\ee
The $1/2$ in front accounts for the fact that we are summing over operator families with even-integer spacing~\cite{Simmons-Duffin:2016wlq}, since we are dealing with identical external fermions. In particular, $P^{4,4}=0$ for even $l$, and $P^{1,1}=P^{1,2}=P^{2,1}=P^{2,2}=P^{3,3}=0$ for odd $l$~\cite{Iliesiu:2015qra}.

After acting with the differential operators we take $1\leftrightarrow 3$ and $(z,\bar{z})\rightarrow (1-z,1-\bar{z})$ and match to individual $s$-channel blocks. We will find the sets $\{i\}$ and $\{j\}$, and the coefficients $A$ and $B$ by matching the $s$-channel. 

There is actually a subtle point we skipped while going from \equref{eq:SomeOverNandL} to \equref{eq:CrossedChannelFinalForm}. As evident from \equref{eq:DifferentialOperators}, we need $S^{\pm\frac{1}{2},\pm\frac{1}{2}}$ to be able to use \equref{eq:SomeOverNandL} for parity-odd structures even though we used $S^{0,0}$ in our ansatz above. We resolve this in appendix~\ref{sec:ExpansionOfS} by expanding $S_a^{r+n,r+m}$ in terms of $S_{a+k}^{r,s}$.

\subsection{Results}
In this section, we first discuss identity matching and find the MFT solutions. Then we consider the exchange of parity-even and parity-odd operators of arbitrary dimension and spin, and calculate their contribution to the OPE coefficients and anomalous dimensions of the double-twist families $[\psi_{(\alpha}\psi_{\beta)}]_{0,l}$, $[\psi_\alpha\psi^\alpha]_{0,l}$, and $[\psi_{(\rho}\partial^\rho_{\;\alpha}\psi_{\beta)}]_{0,l}$ at leading and sub-leading order in the small $z$ expansion. As special cases, we will present the contributions due to stress tensor exchange and scalar exchanges.

The reader is reminded that the contributions of all double-twist families are present, but we simply match the subset of terms relevant for the above families. For example, we match the $\order{z^{1-\Delta_\psi}(1-\bar{z})^0}$ contribution of the stress tensor without matching the $\order{z^{-\Delta_\psi}(1-\bar{z})^1}$ contribution from identity exchange, even though the latter is more dominant in the lightcone limit. However, these contributions come from different twist families in the crossed channel, and there is no mixing for the terms leading order in $(1-\bar{z})$. For instance, both $[\psi_{(\a} \psi_{\b)}]_{0,l}$ and $[\psi_{(\a} \psi_{\b)}]_{1,l}$ bring contributions of order $\order{z^{-\Delta_\psi}(1-\bar{z})^1}$, however only $[\psi_{(\a} \psi_{\b)}]_{0,l}$ brings $\order{z^{-\Delta_\psi}(1-\bar{z})^0}$ terms. So by requiring $[\psi_{(\a} \psi_{\b)}]_{0,l}$ to reproduce these in the crossed channel, we can extract its OPE coefficients and anomalous dimensions.

Below, we will suppress the label for the double twist families whenever there is no ambiguity. For example, we will simply write the OPE coefficient $f^1_{[\psi_{(\a}\psi_{\b)}]_{0,l}}$ as $f^1$. We can extract the relevant family due to the conditions listed in Table~\ref{table:DoubleTwistFamilies} and the fact that the four types of 3-point functions $f^a$ are associated with (parity, spin) as: ($+$, even), ($+$, even), ($-$, even), and ($-$, odd), respectively.

\subsubsection{Identity matching}
Let us first focus on the identity contribution alone. In the $s$-channel the relevant terms are  $\order{z^{-\frac{1}{2}-\Delta_\psi}(1-\bar{z})^0}$ and $\order{z^{\frac{1}{2}-\Delta_\psi}(1-\bar{z})^0}$. We reproduce these terms in the $t$-channel by tuning $A_{a,b,i}$ in \equref{eq:Definition of coefficient A}; for example, we need
\be 
\left\{A_{2,2,\frac{3}{2}-\Delta_\psi},A_{2,2,\frac{5}{2}-\Delta_\psi }\right\}
\ee 
for the double-twist family $[\psi_{(\a} \psi_{\b)} ]_{0,l}$. The fact that $A_{1,1,i}$, $A_{1,2,i}$, and $A_{2,1,i}$ are zero reflects the vanishing of the 3-point coefficient $f^1$ at all orders.

As there is no anomalous dimension for the identity exchange alone we have $\frac{\partial \hb}{\partial l}=1$, hence we can immediately get $P^{a,b}$ with \equref{eq:OPEAnsatz}, then solve for the physical 3-point coefficients\footnote{The $(-1)^l$ term here would be absent in the notation of~\cite{Iliesiu:2015qra}, however we need it as our conformal block normalization in \equref{2.39 of Dolan:2011dv} differs from that paper by a factor of $(-1)^l$.}
\be 
P_\cO^{a,b}=(-1)^lf_{\psi_1\psi_2\cO}^a f_{\psi_3\psi_4\cO}^b\;,
\ee 
where the identity itself has the OPE coefficients $f^1_{\psi\psi\mathbb{1}}=i$ and $f^2_{\psi\psi\mathbb{1}}=0$.

Using the steps described above, we compute the OPE coefficients
\bea[eq:identity OPE]
f^1=\,&0\;, \label{eq: identity lambda 1}
\\
f^2=\,&\frac{f_0}{4\hb^2}\left(1-\frac{8 \Delta_{\psi }-17}{16 \bar{h}}-\frac{256 \Delta_\psi^3-2112 \Delta_\psi ^2+4208 \Delta_\psi-2787}{1536 \hb^2}+\order{\frac{1}{\hb}}^3\right)\;,\\
f^3=\,&\frac{f_0}{2\hb}\sqrt{\frac{\Delta _{\psi }-1}{2\hb}}\left(1-\frac{24\Delta_{\psi }-31}{16 \bar{h}}+\order{\frac{1}{\hb}}^2\right)\;
,\\
f^4=\,&\frac{f_0}{2\hb}\sqrt{\frac{\Delta _{\psi }-1}{2\hb}}\left(1+\frac{8\Delta_{\psi }-1}{16\bar{h}}+\order{\frac{1}{\hb}}^2\right)\;,
\eea
where for convenience we have defined the prefactor $f_0$ as 
\be 
f_0\equiv i \sqrt[4]{\pi } \frac{2^{\frac{3}{2}-\bar{h}} \bar{h}^{\Delta _{\psi }-\frac{1}{4}}}{\Gamma \left(\Delta _{\psi }+\frac{1}{2}\right)}\;.\label{eq: Definition of lambda zero}
\ee 

Note that the results take a slightly simpler form in the $\cD$ basis. E.g., at leading order
\be 
f^2\evaluated_{\cD}=f_0\;,\qquad 
f^{3,4}\evaluated_{\cD}=f_0\sqrt{\frac{\Delta _{\psi }-1}{2\hb}}\;,
\ee 
which follows from \equref{changebasis}.

Using techniques such as~\cite{Karateev:2018oml}, one can actually calculate the full MFT coefficients. We will report on their derivation in a future publication~\cite{spinningInversion}. In the $\D$ basis, they read as
\bea[eq: MFT results]
P_{[\psi_{(\a}\psi_{\b )}]_{0,l}}^{1,1}=\,&P_{[\psi_{(\a}\psi_{\b )}]_{0,l}}^{1,2}=P_{[\psi_{(\a}\psi_{\b )}]_{0,l}}^{2,1}=0\,, \label{eq: MFT results lambda 1}
\\
P_{[\psi_{(\a}\psi_{\b )}]_{0,l}}^{2,2}=\,&\left(f^2\right)^2=-\frac{\sqrt{\pi } \Gamma \left(l+\Delta_\psi -\frac{1}{2}\right)
	\Gamma (l+2 \Delta_\psi -1)}{2^{2 \Delta_\psi +2 l-4}\Gamma \left(\Delta_\psi +\frac{1}{2}\right)^2 \Gamma (l)
	\Gamma (l+\Delta_\psi -1)}\,,
\\
P^{3,3}_{[\psi_{\a}\psi^{\a }]_{0,l}}=\,&\left(f^3\right)^2=
-\frac{\sqrt{\pi } (\Delta_\psi -1) (2 \Delta_\psi +l-1) \Gamma
	\left(l+\Delta_\psi +\frac{1}{2}\right) \Gamma (l+2 \Delta_\psi -2)}{2^{2 \Delta_\psi +2 l-2}\Gamma \left(\Delta_\psi +\frac{1}{2}\right)^2 \Gamma (l+1) \Gamma (l+\Delta_\psi )}\,,
\\
P^{4,4}_{[\psi_{(\rho}\partial^{\rho}_{\;\a}\psi_{\b )}]_{0,l}}\,=&-\left(f^4\right)^2=
\frac{\sqrt{\pi } (\Delta_\psi -1) l \Gamma \left(l+\Delta_\psi
	+\frac{1}{2}\right) \Gamma (l+2 \Delta_\psi -1)}{2^{2 \Delta_\psi +2 l-2}\Gamma \left(\Delta_\psi
	+\frac{1}{2}\right)^2 \Gamma (l+2) \Gamma (l+\Delta_\psi )}\,.
\eea
By changing basis and expanding in large spin, we can reproduce \equref{eq:identity OPE}. We also see that \equref{eq: identity lambda 1} is actually true to all orders, that is $\lambda^1=0$ as we can see from \equref{eq: MFT results lambda 1}. This feature is specific to the leading $n=0$ tower; it is nonzero for $n>0$~\cite{spinningInversion}.

\subsubsection{Matching the exchange of a generic parity-even operator}
Let us turn to the contribution of the exchange of a generic parity-even operator $\cO^+_{\tau,l}$ of twist $\tau$ and spin $l$ in the $s$-channel to the double-twist families $[\psi_{(\alpha}\psi_{\beta)}]_0$, $[\psi_\alpha\psi^\alpha]_0$, and $[\psi_{(\rho}\partial^\rho_{\a}\psi_{\b )}]_0$ in the $t$-channel.

When calculating corrections to the anomalous dimensions of double-twist families we need to recall that the contributions of multiple operators are not additive. Additionally, in general there can be multiple double-twist families that mix with each other. The full formula for their anomalous dimension matrix is\footnote{The anomalous dimension matrix~(\ref{eq:anomalous dimension calculation}) is diagonal if there is no mixing between the double-twist families.}
\be
\frac{\gamma_{[\psi\psi]}}{2}=\delta h_{[\psi\psi]}=\frac{\sum\limits_{\cO}\left(\delta h_{[\psi\psi]}P_{[\psi\psi]}^{a,b}J^{-1}_{[\psi\psi]}\right)_{\cO}}{\sum\limits_{\cO}\left(P_{[\psi\psi]}^{a,b}J^{-1}_{[\psi\psi]}\right)_{\cO}}\;,\label{eq:anomalous dimension calculation}
\ee 
where $J$ is the Jacobian
\be 
J_{[\psi\psi]}\equiv\frac{\partial \delta h_{[\psi\psi]}}{\partial \hb}\;.
\ee 
Here $\cO$ runs over all exchanged operators in the $s$-channel. Likewise, the OPE coefficients $f^a_{[\psi\psi]}$ are given as
\be
(-1)^{\ell}f_{[\psi\psi]}^af_{[\psi\psi]}^b=J_{[\psi\psi]}\sum\limits_{\cO}\left(P^{a,b}_{[\psi\psi]}J^{-1}_{[\psi\psi]}\right)_\cO\;.
\ee 

In the large $\bar{h}$ expansion, we can of course truncate the summation over operators in twist to extract the large $\bar{h}$ behavior. For example, in the first few orders, we see that
\bea[eq:LeadingOrder]
\frac{\gamma_{[\psi\psi]}}{2}=\;\delta h_{[\psi\psi]}
&=\frac{\left(\delta h_{[\psi\psi]}P_{[\psi\psi]}^{a,b}\right)_{\mathbbm{O}}}{\left(P_{[\psi\psi]}^{a,b}\right)_{\mathbb{1}}+\left(P_{[\psi\psi]}^{a,b}\right)_{\mathbbm{O}}}\;,\\
(-1)^lf_{[\psi\psi]}^af_{[\psi\psi]}^b&=\left(P^{a,b}_{[\psi\psi]}\right)_\mathbb{1}+\left(P^{a,b}_{[\psi\psi]}\right)_\mathbbm{O}\;,
\eea
for the identity operator $\mathbb{1}$ along with the operator with minimum twist $\mathbbm{O}$, which is usually either the stress tensor or a scalar of low dimension. 

At leading order in $1/\bar{h}$ one can easily isolate the contribution of any operator. Only the identity operator contributes in the denominator in \equref{eq:anomalous dimension calculation}, allowing one to write an isolated contribution to the anomalous dimension. Likewise, at leading order, one can immediately calculate an individual contribution to $f^a$.

Once we go beyond leading order, we can work with $\left(P_{[\psi\psi]}^{a,b}J^{-1}_{[\psi\psi]}\right)_\cO$ and $\left(\delta h^{a,b}_{[\psi\psi]}P_{[\psi\psi]}^{a,b}J^{-1}_{[\psi\psi]}\right)_\cO$, make an ansatz for their large $\bar{h}$ behavior, and then calculate the corrections to the 3-point coefficients and anomalous dimensions. We find
\begin{subequations}
\label{eq: generic parity-even exchange}
\footnotesize
\be
\left(P^{1,1}J^{-1}\right)_{\cO_{\tau,l}^+}
=&-
\frac{f_+^2}{16 \bar{h}^4 (f_{\cO}'^1)^2 (H_{l+\frac{\tau }{2}-1})^2}\\&\x\left[
\left\{\left(f_{\cO}'^1\right)^2 (5-4 \Delta_\psi )-4 \left(f_{\cO}^2\right)^2 l (l+\tau -1) (-2 \Delta _\psi+\tau +1)^2\right\}^2+\order{\frac{1}{\hb}}
\right]\;,
\ee
\be 
\hspace{-2in}\left(P^{2,2}J^{-1}\right)_{\cO_{\tau,l}^+}
=&-\frac{f_+^2 \left(f_{\cO}'^1\right)^2  }{16 \bar{h}^4}\Bigg[
1+\frac{-8 \Delta _{\psi }+4
	\tau +17}{8 \bar{h}}
+\order{\frac{1}{\hb}}^2\Bigg]\;,
\ee 
\be 
\hspace{.07in}\left((\delta h P)^{2,2}J^{-1}\right)_{\cO_{\tau,l}^+}=&-\frac{f_+^2}{32\hb^4 H_{l+\frac{\tau }{2}-1}}\Bigg[
\left(f_{\cO}'^1\right)^2\left(1+\frac{17+4\tau-8\Delta_\psi}{8\hb}\right)\\&
+\frac{1}{(2\ell-1)(2\ell+2\tau-1)\bar{h}^2} \Bigg( \frac{\left(f_{\cO}'^1\right)^2}{384 } - f_{\cO}'^1f_{\cO}^2 l (\tau -1) (l+\tau -1) \left(-2 \Delta _{\psi }+\tau +1\right)^2
\\&
+\left(f_{\cO}^2\right)^2 l (l+\tau -1) \left(4 l \tau +4 (l-1) l+2 \tau ^2-5 \tau +2\right) \left(2 \Delta _{\psi }-\tau -1\right)^2 \Bigg)\\&
\x\bigg\{
4 l^2 \left(-8 \tau ^3+72 \tau ^2+512 \tau +1851\right)+16 \tau ^4-128 \tau ^3-928 \tau ^2-3214 \tau\\ 
&+1827-4 l \left(8 \tau ^4-80 \tau ^3-440 \tau ^2-1339 \tau +1851\right) -128 (2 l-1) \Delta _{\psi }^3 (2 l+2 \tau -1)\\
&
-16 \Delta _{\psi } \left(l^2 (96 \tau +652)+l \left(96 \tau ^2+556 \tau -652\right)-42 \tau ^2-302 \tau +157\right)\\
&+192 \Delta _{\psi }^2 \left(2 l^2 (\tau +13)+2 l \left(\tau ^2+12 \tau -13\right)-\tau ^2-12 \tau +6\right)
\bigg\}
+\order{\frac{1}{\hb}}^3
\Bigg]\;,
\ee 
\be 
\left(P^{3,3}J^{-1}\right)_{\cO_{\tau,l}^+}=&-\left(P^{4,4}J^{-1}\right)_{\cO_{\tau,l}^+}
\\=&\frac{f_+^2f_{\cO}'^1 }{16\hb^3}\Bigg[
\left(1 + \frac{-8 \Delta _{\psi }+4 \tau +15}{8 \bar{h}}\right) \\
&\times \bigg\{f_{\cO}'^1 (\tau +2)-2 \Delta _{\psi } (f_{\cO}'^1-4 f_{\cO}^2 l (l+\tau -1)) -4 f_{\cO}^2 l (\tau +1) (l+\tau -1)
\bigg\}
+
\order{\frac{1}{\hb}}^2
\Bigg]\;,
\ee 
\be 
\left( (\delta h P)^{3,3}J^{-1}\right)_{\cO_{\tau,l}^+}
=& -\frac{f_+^2}{32 \bar{h}^3 H_{l+\frac{\tau }{2}-1} }\Bigg[
f_{\cO}'^1 \left( f_{\cO}'^1(2\Delta_{\psi}-\tau-2) -4 f_{\cO}^2 \ell(2\Delta_{\psi}-\tau-1)(\ell+\tau-1) \right)\\
&- \frac{1}{8\bar{h}} \Bigg( (f_{\cO}'^1)^2 \left(48 \Delta_{\psi}^2 + \tau (4\tau+35) - 2 \Delta_{\psi}(16\tau+55)+62 \right) \\
&-4 f_{\cO}'^1 f_{\cO}^2 \ell (24 \Delta_{\psi}-4\tau-23)(2\Delta_{\psi}-\tau-1)(\ell+\tau-1) \\
&+16 (f_{\cO}^2)^2 \ell (\tau-2)(\ell+\tau-1) (2\Delta_{\psi}-\tau-1)^2
\Bigg)
+\order{\frac{1}{\hb}}^2
\Bigg]\;,
\ee
\be
\left( (\delta hP)^{4,4}J^{-1}\right)_{\cO_{\tau,l}^+}
=& - \left( (\delta hP)^{3,3}J^{-1}\right)_{\cO_{\tau,l}^+} +\frac{f_+^2}{32 \bar{h}^4 H_{l+\frac{\tau }{2}-1} }\Bigg[(f_{\cO}'^1)^2 ( 4(\Delta_{\psi}-1)(2\Delta_{\psi}-\tau-2) -\tau)  \\&- 8 f_{\cO}'^1 f_{\cO}^2 \ell (2\Delta_{\psi}-1)(2\Delta_{\psi}-\tau-1)(\ell+\tau-1) \\&+ 4 (f_{\cO}^2)^2 \ell (\tau-2)(\ell+\tau-1)(2\Delta_{\psi}-\tau-1)^2 +\order{\frac{1}{\hb}}  \Bigg]\;,
\ee
\normalsize
\end{subequations}
where $H_a$ is a Harmonic number, and we defined
\be 
f_{\cO}'^1\equiv f_{\cO}^1+\left(2\Delta_{\psi}-\tau-1\right)\left(2\Delta_{\psi}+\tau-4\right)f_{\cO}^2
\ee 
and
\be 
f_+\equiv \frac{2^{\frac{3}{2} -\bar{h}+ l + \frac{\tau}{2}} \bar{h}^{\Delta _{\psi }-\frac{1}{4} (1+2 \tau )}}{\Gamma \left(\Delta _{\psi }+\frac{1-\tau }{2}\right)}\sqrt{\left(l+\frac{\tau }{2}\right)_{\frac{1}{2}}H_{l+\frac{\tau }{2}-1}}
\ee 
for convenience. Note that $f_+$ reduces back to \equref{eq: Definition of lambda zero} for identity exchange after setting $l=\tau=\delta h=0$. The appearance of $(f'^1_{\cO})^2$ in the denominator of $P^{1,1}$ is because we compute it by first matching $P^{1,2}$ and then using the relation $P^{1,1} = (P^{1,2})^2/ P^{2,2}$.

\paragraph{Reproducing identity matching:}
As a consistency check, let us use this general form to reproduce the identity exchange contribution to the $[\psi_{(\a}\psi_{\b)}]_{0,l}$ family. We can first check that the anomalous dimension due to sole identity exchange is indeed zero:\footnote{One sees a possible divergence if one does not take $f_{\cO}^2 \rightarrow 0$ first. This is natural because the corresponding 3-point structure does not exist for scalars so its coefficient should be taken to vanish before setting $l = 0$.}
\be
\gamma_{[\psi_{(\a}\psi_{\b)}]_{0,l}}=2\delta h_{[\psi_{(\a}\psi_{\b)}]_{0,l}}=2\frac{\left((\delta h P)^{2,2}\right)_{\cO_{\tau,l}^+}}{\left(P^{2,2}\right)_{\cO_{\tau,l}^+}}\evaluated_{\substack{f_{\cO}^1\rightarrow i,\;f_{\cO}^2\rightarrow 0,\;l\rightarrow 0,\;\tau\rightarrow0}}=0\;.
\ee 

With this, one can now straightforwardly calculate
\bea
\left(f^1\right)^2&=\left(P^{1,1}\right)_{\cO_{\tau,l}^+}\evaluated_{\substack{f_{\cO}^1\rightarrow i,\;f_{\cO}^2\rightarrow 0,\;l\rightarrow 0,\;\tau\rightarrow0}}\;,
\\
\left(f^2\right)^2&=\left(P^{2,2}\right)_{\cO_{\tau,l}^+}\evaluated_{\substack{f_{\cO}^1\rightarrow i,\;f_{\cO}^2\rightarrow 0,\;l\rightarrow 0,\;\tau\rightarrow0}}\;,
\eea
which match \equref{eq:identity OPE}.

\paragraph{Parity-even scalar exchange:} We can also consider the special case of exchange of a parity-even scalar of twist $\tau_s$. Let us write down the anomalous dimension to leading order in $1/\hb$ for convenience. We then only need to use $P^{i,j}$ for identity exchange and $(\delta h P)^{i,j}$ for scalar exchange.

We can immediately read the leading-order anomalous dimensions due to a parity-even scalar exchange as follows:
\begin{subequations}
	\begin{equation}
		\gamma_{[\psi_{(\a}\psi_{\b)}]_{0}}=2\frac{\left((\delta h P)^{2,2}\right)_{\cO_{\tau,l}^+}\evaluated_{\substack{f_{\cO}^1\rightarrow f_\phi,\;f_{\cO}^2\rightarrow 0,\;l\rightarrow 0,\;\tau\rightarrow\tau_s}}}{\left(P^{2,2}\right)_{\cO_{\tau,l}^+}\evaluated_{\substack{f_{\cO}^1\rightarrow f_\mathbb{1},\;f_{\cO}^2\rightarrow 0,\;l\rightarrow 0,\;\tau\rightarrow 0}}}=
		\frac{f ^2_\phi}{\hb^{\tau_s}}\frac{2^{{\tau_s} } \left(\frac{{\tau_s} }{2}\right)_{\frac{1}{2}}  \left(\left(-\frac{{\tau_s} }{2}+\Delta _{\psi }+\frac{1}{2}\right)_{\frac{{\tau_s} }{2}}\right)^2}{\sqrt{\pi }}\;,\label{eq: anomalous of parity-even scalar}
	\end{equation}
	and likewise
	\begin{equation}
		\gamma_{[\psi_\a\psi^\a]_0}=\gamma_{[\psi_{(\rho}\partial^\rho_{\a}\psi_{\b )}]_0}=\frac{f ^2_\phi}{\hb^{\tau_s}}
		\frac{2^{{\tau_s} } \left(\frac{{\tau_s} }{2}\right)_{\frac{1}{2}}  \left(\left(-\frac{{\tau_s} }{2}+\Delta _{\psi }+\frac{1}{2}\right)_{\frac{{\tau_s}
				}{2}}\right)^2\left(\Delta _{\psi }-1-\frac{{\tau_s}}{2}\right)}{\sqrt{\pi } \left(\Delta _{\psi }-1\right)}\;.
	\end{equation}
	\label{eq:AnomalousDimensionOfParityEvenScalars}
\end{subequations}

There are two comments in order. Firstly, $\gamma_{[\psi_\a\psi^\a]_0}$ and $\gamma_{[\psi_{(\rho}\partial^\rho_{\a}\psi_{\b )}]_0}$ na\"ively seem to be divergent in the MFT limit $\Delta_\psi\rightarrow 1$. However, what really matters when solving the analytic bootstrap is the weighted contribution $(f_{{[\psi \psi]}_{0}})^{2} \gamma_{{[\psi \psi]}_{0}}$. We can check from \equref{eq:identity OPE} that the squared OPE coefficients also vanish linearly as $\Delta_{\psi}\rightarrow1$. So, we need to check that this weighted contribution in fact vanishes in mean field theory, or that the factor
\be
\left(-\frac{{\tau_s} }{2}+\Delta _{\psi }+\frac{1}{2}\right)_{\frac{{\tau_s}}{2}} &=& \frac{\Gamma(\Delta _{\psi }+\frac{1}{2})}{\Gamma(-\frac{{\tau_s} }{2}+\Delta _{\psi }+\frac{1}{2})}
\ee 
is zero. In mean field theory we have the contributions from scalar operators $[\psi_{\alpha}\partial^{\alpha\beta}\psi_{\beta}]_{n}$ with twist $2\Delta_{\psi}+1+2n$ for $n\in \N_0$. Plugging in the mean field theory twist we obtain a factor of $\Gamma(-n)^{-1}$, so the result vanishes as expected.

Secondly, we recall that $f_\phi\equiv f_{\psi\psi\phi}$ is purely imaginary due to the Grassmann nature of fermions, hence $\gamma_{[\psi_{(\a}\psi_{\b)}]_0}<0$, as expected for the leading double-twist trajectory~\cite{Komargodski:2012ek}. 

\paragraph{Stress tensor exchange:} As a last example we consider stress tensor exchange. From Ward identities, we know that\footnote{See \cite{Iliesiu:2015qra} for their calculation in the $\cD$ basis.}
\be 
f_{\psi\psi T}^1= \frac{3 i \left(\Delta _{\psi }-1\right) \left(2 \Delta _{\psi }+1\right)}{16 \sqrt{C_T}}\;,\quad 
f_{\psi\psi T}^2= -\frac{3 i}{32 \sqrt{C_T}}\;,
\ee 
where $C_T$ is the central charge. Hence, we compute
\begin{subequations}
	\begin{equation}
\gamma_{[\psi_{(\a}\psi_{\b)}]_{0}}=2\frac{\left((\delta h P)^{2,2}\right)_{\cO_{\tau,l}^+}\evaluated_{\substack{f_{\cO}^1\rightarrow f_{\psi\psi T}^1,\;f_{\cO}^2\rightarrow f_{\psi\psi T}^2,\;l\rightarrow 2,\;\tau\rightarrow 1}}}{\left(P^{2,2}\right)_{\cO_{\tau,l}^+}\evaluated_{\substack{f_{\cO}^1\rightarrow f_\mathbb{1},\;f_{\cO}^2\rightarrow 0,\;l\rightarrow 0,\;\tau\rightarrow 0}}}=
-\frac{48 \Gamma \left(\Delta _{\psi }+\frac{1}{2}\right)^2}{\pi C_T \Gamma \left(\Delta _{\psi }-1\right)^2\hb}\;,
\end{equation}
and
\begin{equation}
\gamma_{[\psi_\a\psi^\a]_0}=\gamma_{[\psi_{(\rho}\partial^\rho_{\a}\psi_{\b )}]_0}= -\frac{24 \left(\Delta _{\psi }-1\right) \left(2 \Delta _{\psi }+1\right) \Gamma \left(\Delta _{\psi }+\frac{1}{2}\right)^2}{\pi C_T \Gamma \left(\Delta _{\psi }\right)^2\hb}\;.
\end{equation}
\end{subequations}

\subsubsection{Matching the exchange of a generic parity-odd operator}
Analogously to the previous section, we can also consider the contribution of the exchange of a generic parity-odd operator $\cO^-_{\tau,\ell}$ in the $s$-channel to the double-twist operators $[\psi_{(\alpha}\psi_{\beta)}]_0$, $[\psi_\alpha\psi^\alpha]_0$, and $[\psi_{(\rho}\partial^\rho_{\a}\psi_{\b )}]_0$ in the $t$-channel.

The analogue of \equref{eq: generic parity-even exchange} now reads as
\begin{subequations}
\label{eq: generic parity-odd exchange}
\footnotesize
\be 
\left(P^{1,2}J^{-1}\right)_{\cO_{\tau,l}^-}
=-f_-^2\Bigg[\left(\left(f_\cO'^3\right)^2-\left(f_\cO'^4\right)^2\right)\left(1+\frac{-8 \Delta _{\psi }+4 \tau +13}{8 \bar{h}}\right)+\order{\frac{1}{\hb}}^2\Bigg]\;,
\ee 
\begin{multline}
\left((\delta h P)^{2,2}J^{-1}\right)_{\cO^-_{\tau,l}}
=-\frac{f_-^2}{8\hb^2}\Bigg[\frac{(2 l+\tau -1) \left(\left(f_\cO'^3\right)^2l-\left(f_\cO'^4\right)^2 (\tau+l-1)\right)}{l (l+\tau -1)}\\\x\left(1+\frac{-8 \Delta _{\psi }+4 \tau +21}{8 \bar{h}}\right)+\order{\frac{1}{\hb}}^2\Bigg],
\end{multline}
\be 
\left((\delta h P)^{3,3}J^{-1}\right)_{\cO^-_{\tau,l}}
=
-\frac{f_-^2}{4}&\Bigg[\left(\left(f_\cO'^4\right)^2+\left(f_\cO'^3\right)^2\right) (2 l+\tau -1)+\frac{2l+\tau-1}{8\hb}\\&\x\bigg\{\frac{\left(f_\cO'^4\right)^2 \left(8 (1-3l) \Delta _{\psi }+4 (l-1) \tau +23 l-4\right)}{l}
\\&+\frac{\left(f_\cO'^3\right)^2 \left(-8 \Delta _{\psi } (3l+3\tau-2)+(l+\tau)(4 \tau +23)-19\right)}{l+\tau -1}
\bigg\}+\order{\frac{1}{\hb}}^2\Bigg],
\ee 
\be 
\left((\delta h P)^{4,4}J^{-1}\right)_{\cO^-_{\tau,l}}
=
-\frac{f_-^2}{4}&\Bigg[\left(\left(f_\cO'^4\right)^2+\left(f_\cO'^3\right)^2\right) (2 l+\tau -1)+\frac{2l+\tau-1}{8\hb}\\&\x\bigg\{\frac{\left(f_\cO'^4\right)^2  \left(8 (l-1) \Delta _{\psi }+4 (l+1) \tau +7 l+4\right)}{l}
\\&+\frac{\left(f_\cO'^3\right)^2 \left(8 \Delta _{\psi } (l+\tau)+l(4 \tau +7)+\tau(4\tau-1)-11\right)}{l+\tau -1}
\bigg\}+\order{\frac{1}{\hb}}^2\Bigg],
\ee 
\normalsize
\end{subequations}
where for convenience we have defined
\be 
f_\cO'^3\equiv{}&2(\tau+l-1)f_\cO^3\;,\\
f_\cO'^4\equiv{}&2lf_\cO^4\;,\\
f_-\equiv{}& \frac{2^{\frac{\tau }{2}-\bar{h}+l-1} \bar{h}^{\Delta _{\psi }-\frac{1}{4} (7+2 \tau )}}{\Gamma \left(\Delta _{\psi }-\frac{\tau }{2}\right)\sqrt{\left(l+\frac{\tau }{2}\right)_{\frac{1}{2}}}}\;.
\ee 
Note that $f_\cO'^3$ and $f_\cO'^4$ are actually the 3-point coefficients in the $\cD$ basis as one can see by comparing \equref{eq:DifferentialOperators} and \equref{changebasis}. Quantities not shown, e.g. $P^{1,1}$ and $P^{2,2}$, do not appear in the matching conditions at this order, hence we do not learn about any new contributions to them. Let us also comment that since $P^{1,2}=0$ in the free theory limit, $P^{1,2}$ being required to be nonzero serves as a probe of the effect of interactions.

\paragraph{Parity-odd scalar exchange:} At leading order, the exchange of a parity-odd scalar contributes to the anomalous dimension of the double-twist families as
\begin{subequations}
\label{eq:AnomalousDimensionOfParityOddScalars}
\be 
\gamma_{[\psi_{(\a} \psi_{\b)}]_0}={}2\frac{\left((\delta h P)^{2,2}\right)_{\cO_{\tau,l}^-}\evaluated_{\substack{f_{\cO}'^3\rightarrow f_\phi,\;f_{\cO}^4\rightarrow 0,\;l\rightarrow 0,\tau\rightarrow\tau_s}}}{\left(P^{2,2}\right)_{\cO_{\tau,l}^+}\evaluated_{\substack{f_{\cO}^1\rightarrow f_\mathbb{1},\;f_{\cO}^2\rightarrow 0,\;l\rightarrow 0,\tau\rightarrow 0}}}=\frac{f_\phi^2}{\hb^{\tau_s+1}}
\frac{ 2^{\tau _s-1} \left(\left(\frac{2 \Delta _{\psi }-\tau
		_s}{2}\right)_{\frac{\tau _s+1}{2} }\right)^2}{\sqrt{\pi }\left(\frac{\tau _s}{2}\right)_{\frac{1}{2}}}\;,
\ee 
where $f_\phi$ is given in the more standard $\cD$ basis. Similarly,
\begin{align}
\gamma_{[\psi_\a\psi^\a]_0}= - \gamma_{[\psi_{(\rho}\partial^\rho_{\a}\psi_{\b)}]_0} ={}&\frac{f ^2_\phi}{\hb^{\tau_s}}
\frac{2^{{\tau_s} } \left(\left(\frac{2 \Delta _{\psi }-\tau
		_s}{2}\right)_{\frac{\tau _s+1}{2} }\right)^2}{\sqrt{\pi } \left(\frac{{\tau_s} }{2}\right)_{-\frac{1}{2}}\left(\Delta _{\psi }-1\right)}	\;.
\end{align}
\end{subequations}
 Comparing with \equref{eq:AnomalousDimensionOfParityEvenScalars}, we see that the contribution of parity-odd scalar exchange to the parity-even double-twist family $[\psi_{(\a} \psi_{\b)}]_0$ comes at the higher order $\hb^{-\tau_s-1}$ instead of $\hb^{-\tau_s}$.

\section{Discussion}
\label{sec:discussion}

We see a number of directions to pursue. First, we would like to upgrade the lightcone bootstrap computations for external fermions and other spinning operators to include nonperturbative effects. Concretely, this requires further developing the inversion formulae or 6j symbols for spinning operators, which can be done with the application of weight-shifting operators \cite{Karateev:2017jgd}. We are planning to pursue this direction in future work. Our computations can also be straightforwardly extended to higher-twist trajectories. As more and more data from the numerical bootstrap (and other methods) becomes available, it will be interesting to apply this technology to 3d O(N) and Gross-Neveu-Yukawa models, conformal gauge theories in various dimensions, multi-graviton states in holographic CFTs, and more.

Second, the matching that we have considered so far involves relating $\log$ coefficients in a finite number of blocks in the direct channel to the leading shift due to the double-twist anomalous dimensions in the crossed channel. To improve this further it is necessary to include the infinite sum of double-twist operators back into the direct channel. Some technology for doing this using the large-spin asymptotics was introduced in~\cite{Simmons-Duffin:2016wlq,Alday:2016njk}. We hope that the theory of such direct-channel resummations can be developed much further, to include subleading and nonperturbative effects, which are likely needed to reproduce the current and stress tensor at a precision level in the O(2) model. Ultimately one needs to compute infinite sums of products of 6j symbols, which is also closely related to the computation of loop diagrams in AdS.

It is our hope that by understanding how to do analytic bootstrap computations at a precision level, these techniques can eventually be merged with numerical approaches to the bootstrap. For example, can the analytic solutions be used for all spins above a cutoff, with lower spins tackled numerically? Is there an iterative procedure where one takes numerical input, computes some portion of the spectrum analytically, and plugs this back into the numerical bootstrap to improve the result? Can the analytic solutions from the inversion formula be used to obtain a near-optimal basis of functionals for the numerical bootstrap, improving on the analytic functionals being explored in~\cite{Mazac:2018mdx,Mazac:2018ycv}? It will be exciting to find a workable merger along these lines. 

\section*{Acknowledgments}

We thank Eric Perlmutter, David Simmons-Duffin, and Douglas Stanford for relevant discussions.  We also thank the organizers of the Bootstrap 2018 workshop where a portion of this work was completed.
The research of D.P. and S.A. is  supported  by  NSF  grant  PHY-1350180  and  Simons  Foundation  grant  488651 (Simons Collaboration on the Nonperturbative Bootstrap).  
The research of DM is supported by the Walter Burke Institute for Theoretical Physics and the Sherman Fairchild Foundation. This material is based upon work supported by the U.S. Department of Energy, Office of Science, Office of High Energy Physics, under Award Number DE-SC0011632

\appendix

\section{Review of Embedding Formalism}
\label{sec:EmbeddingReview}
We are interested in parity-symmetric CFTs in three spacetime dimensions, and we use the Minkowski metric in mostly plus signature $\eta_{\mu\nu}=\text{diag}(-1,1,1)$. Since the double cover of $\SO(2,1)$ is isomorphic to $\Sp(2,\mathbb{R})$, we can use representations of symplectic group; particularly, the smallest fundamental representation will act on a real two dimensional vector space, describing Majorana fermions. 

As the conformal group $\SO(3,2)$ can be realized linearly in 5d spacetime, we embed $\Sp(2,\mathbb{R})$ representations as the projective null representations of $\Sp(4,\R)$ in this 5d embedding space, use that conformal symmetry acts as linear isometries in this formalism, and project back to physical structures by fixing the extra degrees of freedom in the embedding space.

In practice, one removes the extra degrees of freedom of the embedding space by going to so-called Poincar\'e section,
\be 
X^A \rightarrow (x^\mu,1,x^2)\;,
\ee 
where we are working in the lightcone coordinates $X^A=(X^\mu,X^+,X^-)$ and $X^\pm$ are related to the Cartesian coordinates as $X^\pm=X^4\pm X^3$. One can reverse this projection and lift any point to the embedding space in this Poincar\'e section.\footnote{The only exception is the point $\infty$: one cannot use the same Poincar\'e section $(x^\mu,1,x^2)$ used for finite points for the point at infinity as well. A simple way to see this is as follows. We start with $X^a=(x^\mu,X^+,X^-)$ and impose nullness to obtain $X^A=(x^\mu,X^+,x^2/X^+)$. If we now consider an inversion as $x^\mu_R=x^\mu/x^2$, we see that $X^A_R=(x^\mu/x^2,X^+,(x^2X^+)^{-1})\sim(x^\mu,x^2X^+,1/X^+)$ where we use protectiveness of the representation. For $x^2\ne 0$, we can choose $X^+=1/x^2$ for $X_R^A$ and $X^+=1$ for $X^A$, which means we can write both $x^\mu$ and $x_R^\mu$ using the same Poincar\'e section $(x^\mu,1,x^2)$. For $x^2=0$, this is no longer possible, and the reflected point $x_R^\mu=x^\mu/x^2=\infty$ should be instead in the Poincar\'e section $(x^\mu,x^2,1)$.}

As developed in \cite{Costa:2011mg} and generalized to 3d spinors in \cite{Iliesiu:2015qra}, one can encode spinors by polynomials with auxiliary spinor fields:
\begin{equation}
	\Psi(X,S)\coloneqq S_I \Psi^I(X)\;,\qquad \psi(x,s)\coloneqq s_\alpha \psi^\alpha(x)\;,
\end{equation}
where we follow the conventions of \cite{Iliesiu:2015qra} throughout the paper.

The reduction from embedding space to physical space becomes particularly straightforward in this formalism
\begin{equation}
	\Psi(X,S)=\frac{1}{(X^+)^{\Delta_\psi}}\psi(x,s)\;,
\end{equation}
which works under the identification\footnote{We believe that there is a typo in Eq.~(2.12) of \cite{Iliesiu:2015qra}.}
\begin{equation}
	S_I=\sqrt{X^+}\begin{pmatrix}
		s_\alpha\\ x^\alpha_{\;\;\beta}s^\beta
	\end{pmatrix}\;,
\end{equation}
where we define
\begin{equation}
	X^I_{\;\;J}\coloneqq X^A(\Gamma_A)^I_{\;\;J}\;,\qquad x^\alpha_{\;\;\beta}\coloneqq x^\mu(\gamma_\mu)^\alpha_{\;\;\beta}\;.
\end{equation}
Here $\gamma$ and $\Gamma$ are the gamma matrices for which the generators of $\Sp(2,\mathbb{R})$ and $\Sp(4,\mathbb{R})$ can be defined as $-\frac{i}{4}[\gamma^\mu,\gamma^\nu]$ and $ -\frac{i}{4}[\Gamma^A,\Gamma^B]$, respectively.

As a shorthand notation to describe spinor structures in the embedding space, we define
\be
	\<S_1X_2X_3\dots X_{n-1}S_n\>\coloneqq&-S_1\cdot X_2\cdot X_3\cdots X_{n-1}\cdot S_n\\=& -(S_{1})_I(X_2)^I_{\;\;J}(X_3)^J_{\;\;K}\cdots(X_{n-1})^L_{\;\;M}(S_n)^M.
\ee
The minus sign in the front is a convention choice, which allows us to rewrite the expression above as
\begin{equation}
	\<S_1X_2X_3\dots X_{n-1}S_n\>=(S_{1})_I(X_2)^I_{\;\;J}(X_3)^J_{\;\;K}\cdots(X_{n-1})^L_{\;\;M}\Omega^{MN}(S_n)_N\;,
\end{equation}
which, for example, identifies the two-point functions in embedding and physical spaces
\begin{equation}
	\<\Psi(X_1,S_1)\Psi(X_2,S_2)\>=i\frac{\<S_1S_2\>}{X_{12}^{\Delta_\psi+\frac{1}{2}}}\;,\qquad \<\psi^\alpha(x_1)\psi_\beta(x_2)\>=i\frac{(x_{12})^\alpha_{\;\;\beta}}{x_{12}^{2\Delta_\psi+1}}\;.
\end{equation}

\section{Coefficient Expansions}
\label{sec:ExpansionOfS}
The coefficients $S_a^{r,s}(\hb)$ are defined as
\be
\label{eq:ExpansionCoefficientSGeneralized}
S_a^{r,s}(\hb)\coloneqq\frac{1}{\Gamma(-a-r)\Gamma(-a-s)}\frac{\Gamma(\hb-r)\Gamma(\hb-s)}{\Gamma(2\hb-1)}\frac{\Gamma(\hb-a-1)}{\Gamma(\hb+a+1)}\;,
\ee 
which satisfy the one dimensional Mean Field Theory sum~\cite{Simmons-Duffin:2016wlq}:
\be 
\sum\limits_{\substack{\hb=l-a\\l=0,1,\dots}}S_a^{r,s}(\hb)(1-z)^{\hb} \tfo\left(\hb-r,\hb+s,2\hb,1-z\right)=\frac{z^{r+a}}{(1-z)^a}\;.
\ee	

$S_a^{r,s}(\hb)$ scales like $4^{-\hb}\hb^{-\frac{3}{2}-2a-r-s}$ at large $\hb$, meaning that we can expand $S_a^{r,s}(\hb)$ as
\be 
S_a^{r,s}(\hb)=\sum\limits_{k=0}^\infty c_{k,a}^{r,s,m,n} S_{a-\frac{m+n-k}{2}}^{r+m,s+n}(\hb)\;,
\ee
with $\hb$-independent coefficients $c_{k,a}^{r,s,m,n}$. Since we are working at next-to-next-to-leading order in $1/\hb$, we can truncate this expansion as
\be 
S_a^{r,s}(\hb)\approxeq c_{0,a}^{r,s,m,n} S_{a-\frac{m+n}{2}}^{r+m,s+n}(\hb)+c_{1,a}^{r,s,m,n} S_{a-\frac{m+n-1}{2}}^{r+m,s+n}(\hb)+c_{2,a}^{r,s,m,n} S_{a-\frac{m+n-2}{2}}^{r+m,s+n}(\hb)\;.
\ee

Armed with this, let us consider the following summation which appears repeatedly in \equref{eq:tChannelExpression1} after insertion of the ansatz in \equref{eq:OPEAnsatz}: 
\be
\sum_{l=0}^\infty\frac{\partial \hb}{\partial l}S_a^{0,0}(\hb)(1-z)^{\hb}\tfo\left(\hb-m,\hb+n,2\hb,1-z\right)\;.
\ee
For parity-even structures, $m,n=0$, hence we can immediately use \equref{eq:SomeOverNandL}. For parity-odd structures, we can expand $S^{0,0}$ in terms of $S^{m,n}$ and obtain
\begin{multline}
	\sum_{l=0}^\infty\frac{\partial \hb}{\partial l}S_a^{0,0}(\hb)(1-z)^{\hb}\tfo\left(\hb-m,\hb+n,2\hb,1-z\right)\\
	=z^{a+\frac{m-n}{2}} \left(c_{0,a}^{r,s}+c_{1,a}^{r,s}{\sqrt{z}}+\left(\frac{2a-m-n}{2}c_{0,a}^{r,s}+c_{2,a}^{r,s}\right)z+\order{z^{3/2}}\right)\;,
\end{multline}
where $c_{k,a}^{m,n}\equiv c_{k,a}^{0,0,m,n}$. This equation simply means that we replace factors of $\left(\frac{z}{1-z}\right)^a$ in \equref{eq:CrossedChannelFinalForm} with the expression above for parity-odd structures. 

We list below the coefficients $c_0$, $c_1$, and $c_2$ for the reader's convenience:
\be 
c_{k,a}^{r,s,m,n}=\kappa_k\frac{\Gamma \left(-a-\frac{m}{2}+\frac{n}{2}-r-\frac{k}{2}\right) \Gamma \left(-a+\frac{m}{2}-\frac{n}{2}-s-\frac{k}{2}\right)}{\Gamma (-a-r) \Gamma (-a-s)}\;,\quad k\in \{0,1,2\}\;,
\ee 
with
\bea
\kappa_0=&1\;,\\
\kappa_1=&-\frac{m(m+2r)+n(n+2s)}{2}\;,\\
\kappa_2=&\frac{1}{8}m^4+\frac{4r-1}{8}m^3+\frac{n(1+2n+4s)+4r(r-1)-2(a+1)}{8}m^2\nn\\
& +\frac{n^2(1+4r)+n(8rs-4)-4a(n-2)+4(a^2-r^2+1)}{8}m\nn\\
&
+\frac{n}{8}\left((n+2s)^2(n-1)-2n-2a(n-4)+4(a^2+1)\right)\;.
\eea

%%%%%%%%%%%%%%%%%%%%%% BACK MATTER %%%%%%

	\bibliography{collectiveReferenceLibrary}{}
	\bibliographystyle{utphys}

\end{document}